# Mathematical modeling of microRNA-mediated mechanisms of translation repression

Andrei Zinovyev[1,2,3], Nadya Morozova[4], Alexander Gorban[5], Annick Harel-Belan[4]

[1]*Institut Curie, 26 rue d'Ulm, F-75248 Paris France*

[2]*INSERM, U900, Paris, F-75248 France*

[3]*Mines ParisTech, Fontainebleau, F-77300 France*

[4] *CNRS FRE 3377, Laboratoire Epigenetique et Cancer CEA Saclay, 91191 Gif-sur-Yvette, France*

[5]*University of Leicester, Centre for Mathematical Modelling, Leicester, UK*

**Abstract**

MicroRNAs can affect the protein translation using nine mechanistically different mechanisms, including repression of initiation and degradation of the transcript. There is a hot debate in the current literature about which mechanism and in which situations has a dominant role in living cells. The worst, same experimental systems dealing with the same pairs of mRNA and miRNA can provide ambiguous evidences about which is the actual mechanism of translation repression observed in the experiment. We start with reviewing the current knowledge of various mechanisms of miRNA action and suggest that mathematical modeling can help resolving some of the controversial interpretations. We describe three simple mathematical models of miRNA translation that can be used as tools in interpreting the experimental data on the dynamics of protein synthesis. The most complex model developed by us includes all known mechanisms of miRNA action. It allowed us to study possible dynamical patterns corresponding to different miRNA-mediated mechanisms of translation repression and to suggest concrete recipes on determining the dominant mechanism of miRNA action in the form of kinetic signatures. Using computational experiments and systematizing existing evidences from the literature, we justify a hypothesis about co-existence of distinct miRNA-mediated mechanisms of translation repression. The actually observed mechanism will be that acting on or changing the limiting "place" of the translation process. The limiting place can vary from one experimental setting to another. This model explains the majority of existing controversies reported.

**Introduction**

MicroRNAs (miRNAs) are short (21-23nt long) non coding RNAs that are currently considered as key regulators of a wide variety of biological pathways, including development, differentiation and tumorogenesis. Recently, remarkable progress has been made in understanding miRNA biogenesis, function and mode of action. Mature miRNAs are incorporated into the RISC complex, whose key component is an Argonaute protein, and consequently regulate gene expression by guiding the RISC complex toward specific target mRNAs (see Figure 13.1). However, the exact mechanism of this regulation is still a matter of debate. In the past few years, several possible mechanisms have been documented [5,9-10,13,16,19,21,32,41,57,68]. The most documented mechanisms are negative post-transcriptional regulation of mRNA by mRNA translation inhibition and/or mRNA decay, however, some observations show that miRNAs may also act at the transcriptional level.

There is a big controversy in the current literature about which mechanism and in which situations has a dominant role in living cells. The worst, same experimental systems dealing with the same pairs of mRNA and miRNA can provide controversial evidences about which is the actual mechanism of translation repression observed in the experiment. In this chapter we claim that using mathematical modeling can shed light on resolving contradicting experiment interpretations.

The structure of the chapter is the following:

First, we review the whole corpus of available experimental evidences suggesting existence of various mechanisms of miRNA action. Second, we give a detailed description of three mathematical models all describing the process of protein translation in the presence of miRNA. We start with the simplest linear model, suggested before by Nissan and Parker [62]. By analytical analysis of this simple model we demonstrate the importance of exploiting not only the stationary properties but also the dynamical properties in interpreting the experiments on miRNA-mediated silencing of translation. The second model of protein translation, also suggested first by Nissan and Parker and analyzed in [90] shows how recycling of initiation factors and ribosomal subuntis can be taken explicitly into account and to what limitation effects this can lead. We finalize the chapter by describing a mathematical model in which all nine known mechanisms of miRNA action are taken into account, developed by the authors of this chapter [24, 60]. Based on this model, we formulate practical recipes of distinguishing mechanisms of miRNA action by observing stationary and dynamical properties of three quantities: total amount of mRNA, amount of protein synthesized and the average number of ribosomes located on one transcript.

We analyze all three models following a common recipe. The purpose of the analysis is to obtain understanding of how the stationary states and the relaxation times of the model variables depend on model parameters. Though analysis of the stationary state is a well-known approach, analysis of relaxation time is a relatively poorly explored method in systems biology. By definition, the relaxation time is *the characteristic time needed for a dynamic variable to change from the initial condition to some close vicinity of the stationary state*. The relaxation time is a relatively easily observable quantity, and in some experimental methods it is an essential measurement (relaxometry, for example, see [75]). Most naturally the relaxation time is introduced in the case of a linear relaxation dynamics. For example, if a variable follows simple dynamics in the form $x(t) = A(1 - e^{-\lambda t})$, where $A$ is the steady-state value of $x$, then the relaxation time is $\tau = \frac{1}{\lambda}$ and it is the time needed for $x$ to increase from the zero initial value to approximately $1/e \approx 63\%$ of the $A$ value. Measuring the approximate relaxation time in practical applications consists in fitting the linear dynamics to the experimental time curves and estimating λ (for example, see [27]).

The most complete model allows us to simulate the scenario when several concurrent miRNA mechanisms act at the same time. We show that in this situation interpretation of a biological experiment might be ambiguous and dependent on the context of the experimental settings. This allows us to suggest a hypothesis that most of the controversies published in the literature can be attributed to the fact of co-occurrence of several miRNA mechanisms of action, when the observable mechanism acts on the limiting step of protein translation which can change from one experiment to another.

## 1. Review of published experimental data supporting each of proposed mechanisms of microRNA action

Protein translation is a multistep process which can be represented as sequence of stages (initiation, ribosome assembly, elongation, termination) involving circularization of mRNA, recruiting the mRNA cap structure and several protein initiation factors and ribosomal components. The process of normal translation can be regulated by small non-coding microRNAs through multiple mechanisms (Figure 13.1).

Here we are reviewing available experimental data on all reported mechanisms of microRNA action, grouping them in a way which elucidates the main details supporting each of these proposed mechanisms.

## M1: Cap-40S Initiation Inhibition

Inhibition of cap recruitment as a suggested mechanism of microRNA repression was initially proposed by Pillai, et al [69], and since that time this mechanism was one of the most frequently identified [17,19,29,40,80,91].

The main evidence in favor of the cap-recognition and 40S assembly inhibition model was that IRES-driven or A-capped mRNA (see Figure 13.1 for definitions of these terms) are refractory to microRNA inhibition, together with a shift toward the light fraction in the polysomal gradient. According to this, an initiation mechanism upstream of eIF4G recruitment by eIF4E was postulated and it was hypothesised that it suppresses the recognition of the cap by eIF4E. The very recent studies [17, 91] detailed GW182 involvement in the initiation suppression via cap-40S association, thus providing additional evidence for this mechanism.

## M2: 60S Ribosomal Unit Joining Inhibition

It has also been proposed that microRNA could act in a later step of initiation, i.e., block the 60S subunit joining. This hypothesis, initially suggested by Chendrimada et al. [11], was next supported by in-vitro experiments showing a lower amount of 60S relative to 40S on inhibited mRNAs, while toe-printing experiments show that 40S is positioned on the *AUG* codon [85]. It is important to point out that, strictly speaking, there is no proof that miRNA affects the scanning for the *AUG* codon in this work, although some works interpret this data as an inhibition of scanning [62].

## M3: Elongation Inhibition

Historically, the inhibition of translation elongation mechanism was the first proposed mechanism for microRNA action [64]. The major observation supporting this hypothesis was that the inhibited mRNA remained associated with the polysomal fraction, which was reproduced in different systems [25,45,54,67]. The idea of a post-initiation mechanism was further supported by the observation that some mRNAs can be repressed by microRNA even when their translation is cap-independent (see Figure 13.1, mRNAs with an IRES or A-capped) [4,37,53,67,54].

Actually, in the work by Olsen and Ambros [64] there is no data supporting elongation inhibition rather than other post-initiation mechanisms (e.g. nascent polypeptide degradation), because the main conclusion is derived only by studying the polysomal profiles. But some evidences can be found in the work by Gu et al. [25], describing that on the same mRNA, when the ORF is prolongated downstream the binding site of miRNA (mutation in the stop codon), the inhibition by a miRNA is lost. If a rare (slow) codon is introduced upstream the binding site, the inhibition is relieved, which shows that the presence of actively translating ribosomes on the binding site impairs the inhibition by miRNA. The presence of a normal polysomal distribution of the inhibited mRNA and sensitivity to EDTA (ethylenediaminetetraacetic acid) and puromycin indicating functional, translating polysomes, allowed the authors to suggest the "elongation" model. Also some data of Maroney et al. [54], could also imply that elongation is slowed down by microRNA (as the ribosome stays longer on the inhibited mRNA), but the authors discussed this point critically and were not able to reproduce it in vitro.

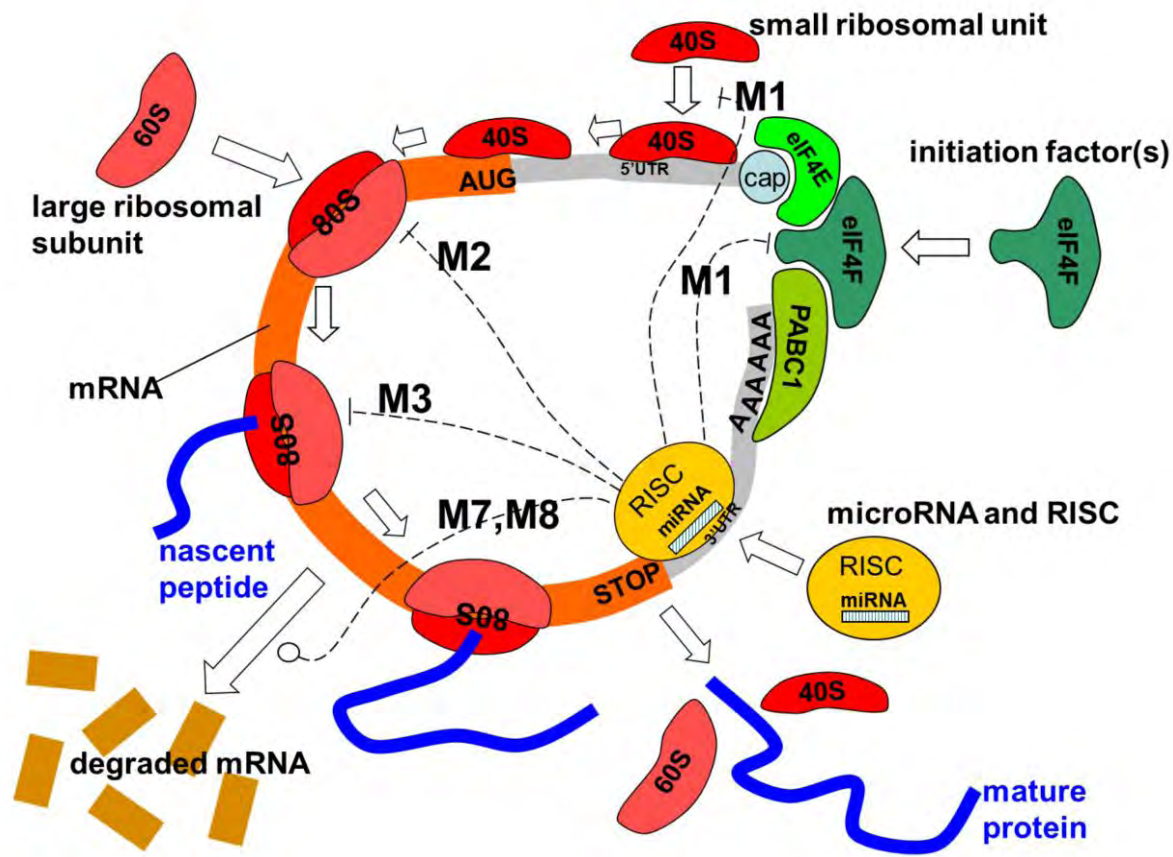

Figure 13.1. Interaction of microRNA with protein translation process. Several (from nine documented) mechanisms of translation repression are shown: M1) on the initiation process, preventing assembling of the initiation complex or recruiting the 40S ribosomal subunit; M2) on the ribosome assembly; M3) on the translation process; M7,M8) on the degradation of mRNA. There exist other mechanisms of microRNA action on protein translation (transcriptional, transport to P-bodies, ribosome drop-off, co-translational protein degradation and others) that are not visualized here. Here 40S and 60S are light and heavy components of the ribosome, 80S is the assembled ribosome bound to mRNA, eIF4F is an translation initiation factor, PABC1 is the Poly-A binding protein, "cap" is the mRNA cap structure needed for mRNA circularization (which can be the normal m7G-cap or artificial modified A-cap). The initiation of mRNA can proceed in a cap-independent manner, through recruiting 40S to IRES (Internal Ribosome Entry Site) located in 5'UTR region. The actual work of RNA silencing is performed by RISC (RNA-induced silencing complex) in which the main catalytic subunit is one of the Argonaute proteins (AGO), and miRNA serves as a template for recognizing specific mRNA sequences.

### M4: Ribosome Drop-off (premature termination)

First (and seems to be the only one till today) evidence of this mechanism was done by Petersen et al. [67], who observed no difference in polysomal profile in the presence of miRNA. Addition of puromycin, which necessitate peptidyl transferase activity to act, didn't change the polysomal profile in the presence or in the absence of the miRNA. The authors have concluded that polysomes are actively translating even in presence of miRNA. They were not be able to detect any peptide by radiolabelling and therefore postulated the ribosome drop-off mechanism.

However, Wang et al. [84] presented data also supporting premature termination: the read-through codon-stop and more rapid loss of polyribosome upon initiation block.

### M5: Co-translational Nascent Protein Degradation

Initially, the idea of nascent protein degradation was proposed by Nottrott et al. [63], according to the presence of inhibited mRNA and AGO protein in polysomes, which suggests the action of miRNA on actively translated mRNA. However, no nascent peptide has ever been experimentally demonstrated [63,67,68,84]; thus the nascent polypeptide degradation, if it exists, should occur extremely rapidly after the synthesis. Anyhow, being able to immunoprecipitate the nascent polypeptide together with the mRNA and the polysome in the case of normal translation, the authors failed to do so in the case of miRNA inhibition [63]. Pillai et al. in [68] showed that this degradation, if exists, should be proteasome-independent, and no specific protease or complex involved in this polypeptide degradation has ever been identified. Data supporting nascent polypeptide degradation are the following: a) sedimentation of mRNA together with miRNA-RISC complexes in actively translating (puromycin-sensitive) polysomes; b) blocking the initiation (in a cap-dependent manner), resulted in a shift in polysomal profile, suggesting that the repressed mRNA is actively transcribed. In Wang et al. [84] the authors also support nascent protein degradation showing polysomal distribution with puromycin sensitivity, but in the same paper they also present data supporting premature termination. Maroney et al. [54] and Gu et al. [25] presented experimental data which are very coherent with this line though not concluding that this mechanism is the dominating one: presence of miRNA/mRNA complex in polysomes, sensitivity to different conditions is an indication of translating ribosomes.

**M6: Sequestration in P-bodies**

An effect of sequestration of mRNA targeted by AGO-microRNA complex in cytoplasmic structures called P-bodies was initially shown in [69,78]. Next this result was confirmed in many studies characterising P-bodies as structures where the translational machinery is absent and the degradation machinery is functional [8,33,47,50,51,66,69,78]. The main propositions about P-bodies' function was that they sequestrate targeted mRNA apart from translational machinery, or that P-bodies give a kinetics advantage for mRNA decay (local concentration of all needed enzymes). Two additional important points were elucidated in [14], showing that mRNA localised in the P-bodies, can be still associated with polysomes, and also that miRNA silencing is still possible when P-bodies are disrupted. This led to the conclusion that P-bodies are not required for but rather a consequence of microRNA-driven RNA degradation or translational inhibition. This conclusion is also supported by other studies [48] and is mostly accepted today. Moreover, only a small portion of miRNA, mRNA and RISC complex is localised inside macroscopic P-bodies [48,69].

**M7: mRNA Decay (destabilisation)**

Recently, starting from the first description by Lim et al. [49], a lot of data has revealed miRNA-mediated mRNA decay (destabilization) of targeted mRNA without direct cleavage at the binding site [2,3,6,15,34,49,77,83,87]. Also, most of the authors note that only a slight protein decrease can be obtained by translational inhibition only. When the protein level decreases by more than 33%, mRNA destabilization is the major component of microRNA-driven silencing [2]. Anyhow, all these data, concordant in the main point (mRNA decay mechanism), are different in details of its concrete mechanism (decay by mRNA deadenylation, decapping, or 5′ to 3′ degradation of the mRNA). In the review of Valencia-Sanchez et al. [81], it is concluded that the decapping followed by 5'->3' degradation is the most plausible mechanism for the miRNA inhibition, while deadenylation could lead only to a decrease in the initiation efficiency by disrupting the loop between polyA and cap. Behn-Ansmant et al. [6], showed that GW182, an AGO partner in the microRNA pathway, triggers deadenilation and decapping of bound mRNA, which leads to mRNA decay. Filipowicz et al. [19] supports the idea of the degradation running mostly in 5'->3' direction after deadenylation and decapping, in contradiction to [87], where it is claimed that deadenylation is the principal cause of the mRNA decay but degradation goes first in the 3'->5' way. The latter work also indicated that the degradation mechanism is supposed to be only an addition to the translational inhibition and that translational inhibition has the same efficiency with or without degradation. Coller and Parker in 2004 proposed that as the poly(A) tail can enhance translation rates and inhibit mRNA decay, then the increases of deadenylation rates by miRNA/RISC could be counted as additional mechanism by which translation repression and mRNA decay could be stimulated [12]. Finally, Eulalio et al. showed that there could be two different cases in mRNA degradation by microRNA: in one the ongoing translation is required for the decay, and in the second the decay occurs in the absence of active translation, and assume that this depends on mRNAs undergoing the decay [15].

**M8: mRNA Cleavage**

mRNA cleavage (similar to what is observed with siRNA) can be observed when the sequence of microRNA is completely or almost completely complementary to its target binding site. The first proposition for this mechanism was made for plants [52,73], and since that time, the miRNA-mediated mRNA cleavage was proved to be very common for plants, and much more rare in animals [31,89].

Though the most of known mammalian microRNAs are only partially complementary to their targets, there is some data on miRNA-mediated mRNA cleavage, for example, for miR-196 [89]. A few other works (e.g., in mammals [1,81] or in *C.elegans* [3]) also mentioned cleavage as a possible mechanism of microRNA repression in animals.

**M9: Transcriptional Inhibition through microRNA-mediated chromatin reorganization following by gene silencing**

Although the first publication for siRNA-mediated transcriptional repression [61] was made in 2004, the first publication proving miRNA-mediated transcriptional repression in mammalian cells appeared only recently [38]. Around this time also appeared the first publication for miRNA-mediated transcriptional activation, showing that microRNA-373 induces expression of genes with complementary promoter sequences [71]. Since then very few evidences of miRNA-directed transcriptional gene silencing (TGS) in mammalian cells were obtained [38,86].

## 2.2. Controversies between the miRNA-mediated mechanisms of translation repression

It is important to note that it is extremely difficult to discriminate experimentally between different potential post-initiation mechanisms, such as elongation inhibition, premature ribosome dissociation ("ribosome drop-off") or normal elongation with nascent polypeptide degradation. Both elongation slowing down and nascent polypeptide degradation are supported by the fact that the mRNA-polysomal association is puromycin-sensitive, indicating polysomes' activity [54,63] and by the observed requirement for microRNA binding in the unstranslated region [25]. Premature ribosome dissociation is supported by decreased read-through of inhibited mRNA [67]. Both ribosome drop-off and ribosomal "slowing down" are supported by the slight decrease in the number of associated ribosomes [54,63]. But, eventually with premature drop-off, the polysomal profile will not be the same as in the case of nascent protein degradation, as one should have less ribosomes per mRNA.

Summarizing overview on the proposed mechanisms, we briefly emphasize the main controversial data.

1. First of all, even the question at which level (transcriptional, translational, etc) the microRNA action takes place is still strongly debated. The most frequently reported, but also very contradictory in details, is the mechanism of gene repression by microRNAs occurring at the level of mRNA translation (this includes mechanisms of arrest at initiation and elongation steps, ribosome drop-off and nascent polypeptide degradation), but repression at the level of mRNA (before translation) have been also proposed as the principal one in many studies (this includes mechanisms of microRNA-mediated mRNA decay, sequestration of target mRNAs in P-bodies and rare in animals but frequent in plants mechanism of target mRNA cleavage). Moreover, it was proposed that some microRNAs mediate chromatin reorganization followed by transcriptional repression, which involves mechanisms strikingly different from the previous modes of repression. Finally, the transcriptional activation by microRNA [38,71] and translational activation by microRNA have been also proposed [65,82].

2. Currently, the action of microRNAs at the level of initiation of translation seems to be the most favourite one accordingly to many recent publications. Anyhow, the experimental data, supporting this mechanism, are also controversial in the result interpretations of different groups suggesting this mechanism. For example, it has been proposed that AGO2 protein could interact with the cap via de eIF4E-like domain and therefore compete with eIF4E for binding the cap [40]. However, this has been weakened by the recent finding that this domain could be involved in the binding with GW182, an important protein for miRNA action, and by crystallographic analysis showing that the folding will not allow such a interaction with the cap [17,39].

The main observation supporting the initiation mechanism is that mRNA with IRES or A-cap can't be inhibited by microRNA, but in the considerable number of works it was shown that some mRNAs can be repressed by a microRNA even when their translation is cap-independent [4,37,53,67].

3. For blocking the 60S subunit joining mechanism, it was shown that eIF6, an inhibitor of 60S joining, is required for microRNA action [11], but this was contradicted by other studies [17].

4. An interesting observation was reported in [42] about that the same mRNA targeted by the same microRNA can be regulated either at the initiation or the elongation step depending on the mRNA promoter. But next, using the same promoter described in [42], as leading to the initiation mechanism, the authors suggests the "elongation" model, according to the polysomal distribution on the inhibited mRNA [25].

5. Different results about mechanisms of microRNA action were obtained depending on the transfection method of the inhibited mRNA [53].

6. Karaa et al. describes the VEGF gene, which is endogenously regulated by a miRNA, miR16, acting on an IRES (see Figure 13.1 legend) [37]. VEGF is translated from one of two IRES, and only one of these IRES allows inhibition by miR16. Therefore, inhibition by microRNA is possible even in IRES-driven translation, but not for all IRES-driven cases, even if those two IRES have been described as similar.

7. Kozak et al., reviewing different papers about miRNA-mediated inhibition, claimed a lot of experiments to be faulty [43] and reported that only few studies are based on reliable experiments could be considered, namely [11,29,40,55,83,85]. The statement of the author that "other suggested mechanisms are not mentioned here because the speculations greatly exceed the facts" seems to concern [53,63,67,69].Together with this, the author is very critical about interpretations of the IRES experiments.

8. Olsen et al. has described inhibition of elongation step, based on the presence of polysomal distribution [64]. But, actually, there is no additional data supporting elongation inhibition rather than nascent polypeptide degradation, because in both works the main (and different!) conclusion is driven only by studying the polysomal profiles.

9. In several studies it was shown that degradation and translational arrest can be coupled in many systems [15,17,18,69,87], but here the situation is also not completely understood: some mRNAs are repressed mostly at the translational level, others mostly at the stability level (with or without a requirement for concurrent translation inhibition), and some at both levels [1]. In some works it is suggested that microRNA-mediated mRNA decay is a consequence of translational repression, the other group of studies suggests that neither the destabilisation is a consequence of translational arrest, nor the translational repression is a consequence of degradation, but that the two mechanisms are concurrently occurring [16,17,91]. It has been concluded that the relative contributions of translational repression and decay differ depending on the presence or absence of the poly(A) tail [17]. However, in deciding whether the deadenylation is the cause or consequence of silencing, the authors again present controversial data interpretations [79].

*Thus, the experimental data and summarizing conclusions about the mechanism by which microRNA repress mRNA expression are highly controversial, and though arise a question about interrelations between the different mechanisms and their possible concomitant action, do not consider it in the frame of one unique mechanism of microRNA action.*

*Using a series of mathematical models with increasing complexity, we show how mathematical modelling can help in interpreting the experimental results and even suggest some explanations of the ambiguous observations.*

### 3. Modeling notations and assumptions

In this chapter we consider three mathematical models of miRNA action of increasing complexity:

1. *The simplest linear model of protein translation*. This model was first suggested in [62]. It allows distinguishing two types of miRNA-mediated mechanisms: those acting at the very early stage of translation initiation and those acting at a later stage.

2. *Non-linear model of protein translation taking into account recycling of ribosomes and initiation factors*. This model was first suggested in [62]. It allows distinguishing four types of miRNA-mediated mechanisms: acting at the very early stage of initiation, later stage of initiation, ribosome assembly step, elongation and termination (considered together as one step of translation).

3. *General model describing all known mechanisms of miRNA action*. This model was developed by the authors of this chapter [24,60] and includes nine mechanisms of miRNA action. Using this model, we classify the existing mechanisms by their dynamical properties and suggest a tool to distinguish most of them based on experimental data.

Of course, any mathematical model is a significant simplification of biological reality. The first two models, for example, consider only a limited subset of all possible mechanisms of microRNA action on the translation process. All processes of synthesis and degradation of mRNA and microRNA are deliberately neglected in these models. Interaction of microRNA and mRNA is simplified: it is supposed that the concentration of microRNA is abundant with respect to mRNA. Interaction of only one type of microRNA and one type of mRNA is considered (not a mix of several microRNAs). The process of initiation is greatly simplified: all initiation factors are represented by only one molecule which is marked as eIF4F.

Finally, the classical chemical kinetics approach is applied, based on solutions of ordinary differential equations, which assumes sufficient and well-stirred amount of both microRNAs and mRNAs. Another assumption in the modeling is the mass action law assumed for the reaction kinetic rates.

It is important to underline the interpretation of certain chemical species considered in the system. The ribosomal subunits and the initiation factors in the model exist in free and bound forms. Moreover, the ribosomal subunits can be bound to several regions of mRNA (the initiation site, the start codon, the coding part). Importantly, several copies of fully assembled ribosomes can be bound to one mRNA. To model this situation, we have to introduce the following quantification rule for chemical species: amount of ``ribosome bound to mRNA'' means the total number of ribosomes translating proteins, which is not equal to the number of mRNAs with ribosome sitting on them, since one mRNA can hold several translating ribosomes (polyribosome). In this view, mRNAs act as *places* or *catalyzers*, where translation takes place, whereas mRNA itself formally is not consumed in the process of translation, but, of course, can be degraded or synthesized.

Let us introduce notations that will be used throughout the chapter for designation of chemical species:

1. *40S*, free small ribosomal subunit.
2. *60S*, free large ribosomal subunit.
3. *eIF4F*, free initiation factor.
4. *M,* free mRNA (models 1 and 2) and mRNA with free initiation site (model 3).
5. *P,* translated protein.
6. *B,* mRNA located in P-bodies.
7. *F,* state of mRNA when the small ribosomal subunit bound to the initiation site.
8. *A,* state of mRNA when the small ribosomal subunit bound to the start codon.
9. *R,* translating ribosome, located on mRNA.

Square brackets will denote the amounts of the corresponding species. For example, [*M*] will denote the amount of free mRNA in the system.

Note that the notations for the kinetic rate constants are not equivalent in three models. For example, while $k_1$ notifies the kinetic rate of the cap initiation in the models 1 and 2, it has different measure units in linear and non-linear models. Moreover, in the model 3, $k_1$ notifies the rate constant for translation initiation (recruiting 40S subunit) of mRNA already in translation process. Hence, the meaning of $k_i$ constants should be considered differently per each model type.

## 4. Simplest linear model of protein translation

The simplest representation of the translation process has the form of a circular cascade of reactions [62] (see Figure 13.2). The model contains four chemical species *40S*, *F*, *A* and *P* and three chemical reactions.

The catalytic cycle in which the protein is produced is formed by the following reactions:

1. 40S → *F*, Initiation complex assembly (rate *k1*).

2. *F* → *A*, Some late and cap-independent initiation steps, such as scanning the 5'UTR for the start *A* codon recognition (rate *k2*) and 60S ribosomal unit joining.

3. *A* → 40S, combined processes of protein elongation and termination, which leads to production of the protein (rate *k3*), and fall off of the ribosome from mRNA.

The model is described by the following system of equations [62]:

$$\begin{cases} \dfrac{d[40S](t)}{dt} = -k1[40S] + k3[A] \\ \dfrac{d[F](t)}{dt} = k1[40S] - k2[F] \\ \dfrac{d[A](t)}{dt} = k2[F] - k3[A] \\ Psynth(t) = k3[A](t) \end{cases} \quad (1)$$

where $Psynth(t)$ is the rate of protein synthesis.

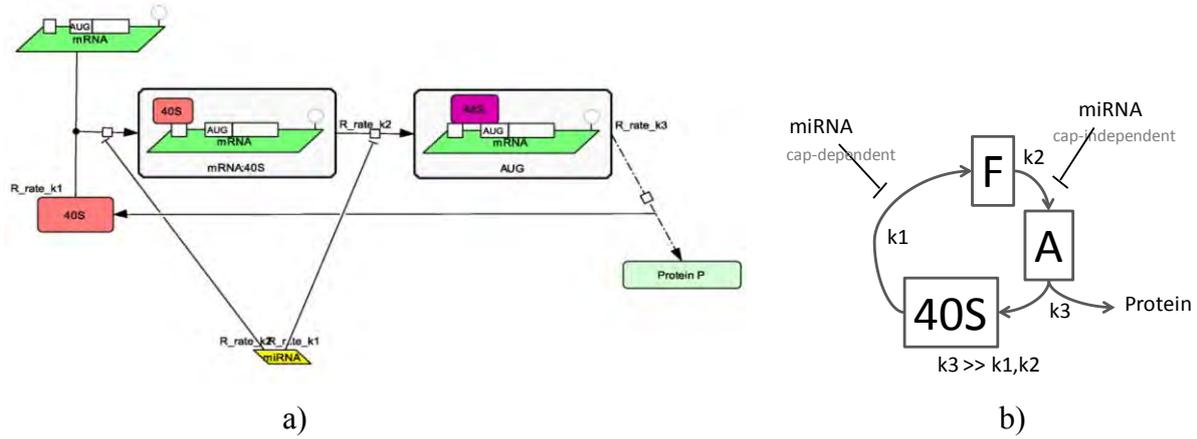

a)                    b)

Figure 13.2. The simplest mathematical model of protein translation which is capable to explain the effect of a miRNA on the very early (rate $k1$) and late (rate $k2$) steps of mRNA initiation; a) graphical presentation of the model in the SBGN standard; b) schematic model presentation. Action of miRNA is modeled by reducing the rate constant of the corresponding translation step.

Following [62], let us assume that $k3 \gg k1, k2$. This choice was justified by the following statement: ``...The subunit joining and protein production rate ($k3$) is faster than $k1$ and $k2$ since $F$ complexes bound to the $A$ without the 60S subunit are generally not observed in translation initiation unless this step is stalled by experimental methods, and elongation is generally thought to not be rate limiting in protein synthesis...'' [62].

Under this condition, the equations (1) have the following approximate solution (which becomes the more exact the smaller the $(k1+k2)/k3$ ratio), suggested earlier in [90]:

$$\begin{bmatrix} 40S(t) \\ F(t) \\ A(t) \end{bmatrix} = \dfrac{40S_0}{\dfrac{1}{k1} + \dfrac{1}{k2}} \left( \begin{bmatrix} 1/k1 \\ 1/k2 \\ 1/k3 \end{bmatrix} + \dfrac{1}{k3}\begin{bmatrix} -1 \\ 1 \\ 0 \end{bmatrix} e^{-k3t} + \dfrac{1}{k2}\begin{bmatrix} 0 \\ 1 \\ -1 \end{bmatrix} e^{-(k1+k2)t} \right), \quad (2)$$

$$Psynth(t) = \dfrac{40S_0}{\dfrac{1}{k1} + \dfrac{1}{k2}} \left( 1 - \dfrac{k3}{k2} e^{-(k1+k2)t} \right) \quad (3)$$

for the initial condition

$$\begin{bmatrix} 40S(t) \\ F(t) \\ A(t) \\ Psynth \end{bmatrix} = \begin{bmatrix} 40S_0 \\ 0 \\ 0 \\ 0 \end{bmatrix}.$$

From the solution (2-3) it follows that the dynamics of the system evolves on two time scales: 1) fast elongation dynamics on the time scale $\approx 1/k3$; and 2) relatively slow translation initiation dynamics with the relaxation time $t_{rel} \approx \frac{1}{k1+k2}$. The protein synthesis rate formula (2-3) does not include the $k3$ rate, since it is neglected with respect to $k1$, $k2$ values. From (2-3) we can extract the formula for the protein synthesis steady-state rate $Psynth(t)$ (multiplier before the parentheses) and the relaxation time $t_{rel}$ for it (inverse of the exponent power):

$$Psynth = \frac{40 S_0}{\frac{1}{k1}+\frac{1}{k2}}, \quad t_{rel} = \frac{1}{k1+k2}. \qquad (4)$$

Now let us consider two experimental situations: 1) the rate constants for the two translation initiation steps are comparable $k1 \approx k2$, and 2) the cap-dependent rate $k1$ is limiting: $k1 << k2$. Accordingly to [62], the second situation can correspond to modified mRNA with an alternative cap-structure (A-cap), which is much less efficient for the assembly of the initiation factors, 40S ribosomal subunit and polyA-binding proteins.

For these two experimental systems (let us call them "wild-type" and "modified" correspondingly), let us study the effect of microRNA action. We will model the microRNA action by diminishing the value of a kinetic rate constant for the reaction representing the step on which the microRNA is acting. Let us assume that there are two alternative mechanisms: 1) microRNA acts in a cap-dependent manner (thus, reducing the $k1$ constant) and 2) microRNA acts in a cap-independent manner, for example, through interfering with 60S subunit joining (thus, reducing the $k2$ constant). The dependence of the steady rate of protein synthesis $Psynth \sim \frac{1}{\frac{1}{k1}+\frac{1}{k2}}$ and the relaxation time $t_{rel} \approx \frac{1}{k1+k2}$ on the efficiency of the microRNA action (i.e., how much it is capable to diminish a rate coefficient) is shown in Figure 13.3.

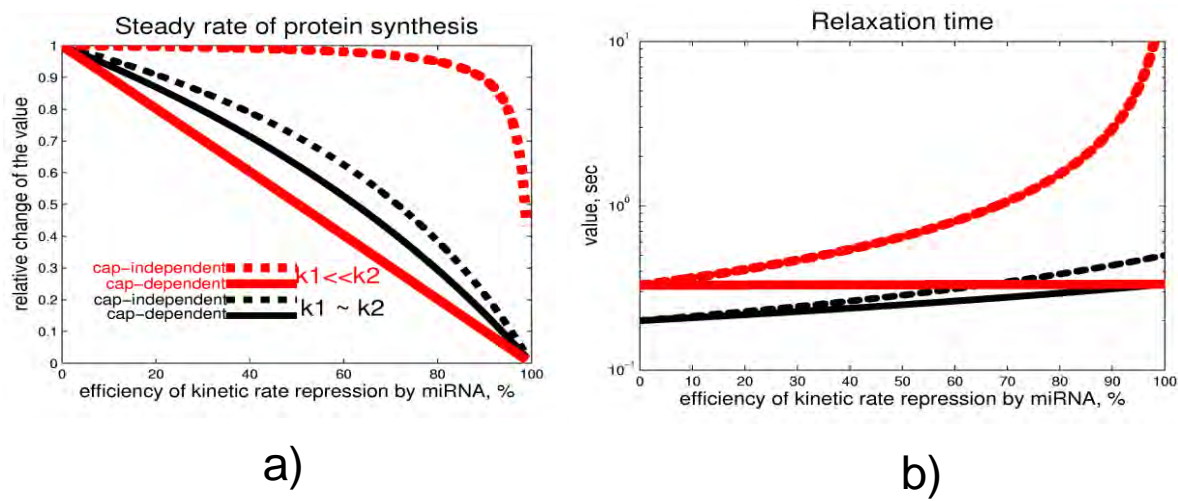

a) b)

Figure 13.3. Dependence of the relative change of the protein synthesis steady rate and the relaxation time (time needed to achieve the steady rate) on the efficiency with which microRNA can act at an early cap-dependent ($k1$) or late cap-independent ($k2$) rate of translation. Two scenarios are considered: a wild-type one when $k1$ value is similar to $k2$ and the case of a modified A-cap structure when $k1 << k2$ even in the absence of miRNA.

Interestingly, experiments with cap structure replacement were made and the effect of microRNA action on the translation was measured [55.80]. No change in the protein rate synthesis after applying microRNA was observed. From this it was concluded that microRNA in this system should act through a cap-dependent mechanism (i.e., the normal ``wild-type" cap is required for microRNA recruitment). It was argued that this could be a misinterpretation [62] since in the "modified" system, cap-dependent translation initiation is a rate limiting process ($k1 << k2$). Hence, even if microRNA acts in the cap-independent manner (inhibiting $k2$), it will have no effect on the final steady state protein synthesis rate. This was confirmed by the graph similar to the Figure 13.3a.

From the analytical solution (2-3) we can further develop this idea and claim that it is possible to detect the action of microRNA in the "modified" system if one measures the protein synthesis relaxation time: if it significantly increases then microRNA probably acts in the cap-independent manner despite the fact that the steady state rate of the protein synthesis does not change. This is a simple consequence of the fact that the relaxation time in a cycle of biochemical reactions is limited by the second slowest reaction, see [22,90]. If the relaxation time does not change in the presence of microRNA then we can conclude that none of the two alternative mechanisms of microRNA-based translation repression is activated in the system, hence, microRNA action is dependent on the structure of the "wild-type" transcript cap.

The observations from the Figure 13.3 are recapitulated in the Table 13.1. This analysis (of course, over-simplified in many aspects) provides us with an important lesson: observed dynamical features of the translation process with and without presence of microRNA can give clues on the mechanisms of microRNA action and help to distinguish them in a particular experimental situation. Theoretical analysis of the translation dynamics highlights the important characteristics of the dynamics which should be measured in order to infer the possible microRNA mechanism.

This conclusion suggests the notion of a **kinetic signature of microRNA action mechanism** which we define as *a set of measurable characteristics of the translational machinery dynamics (features of time series for protein, mRNA, ribosomal subunits concentrations) and the predicted tendencies of their changes as a response to microRNA action through a particular biochemical mechanism.*

Table 13.1. Modeling two mechanisms of microRNA action on several translation steps in the simplest linear model

| Observable value | Initiation | Step after initiation, cap-independent | Elongation |
|---|---|---|---|
| Wild-type cap | | | |
| *Steady-state rate of protein synthesis* | decreases | decreases | no change |
| *Relaxation time of protein synthesis* | increases slightly | increases slightly | no change |
| A-cap | | | |
| *Steady-state rate of protein synthesis* | decreases | no change | no change |
| *Relaxation time of protein synthesis* | no change | increases drastically | no change |

## 5. Non-linear Nissan and Parker's model of protein translation

To explain the effect of microRNA interference with translation initiation factors, a non-linear version of the translation model was proposed in [62] which explicitly takes into account recycling of initiation factors (eIF4F) and ribosomal subunits (40S and 60S).

### 5.1. Model equations and the steady state solutions

The model contains the following list of chemical species (Figure 13.4): *40S*, *60S*, *eIF4F*, *F*, *A*, and *R* and four reactions, all considered to be irreversible:

1. *40S + eIF4F → F*, assembly of the initiation complex (rate *k1*).

2. *F → A*, some late and cap-independent initiation steps, such as scanning the 5'UTR for the start codon *A* (rate *k2*).

3. *A → R*, assembly of ribosomes and protein translation (rate *k3*).

4. 80S → 60S+40S, recycling of ribosomal subunits (rate *k4*).

The model is described by the following system of equations [62]:

$$\begin{cases} \dfrac{d[40S](t)}{dt} = -k1[40S][eIF4F] + k4[R] \\ \dfrac{d[eIF4F]}{dt} = -k1[40S][eIF4F] + k2[F] \\ \dfrac{d[F](t)}{dt} = k1[40S][eIF4F] - k2[F] \\ \dfrac{d[A](t)}{dt} = k2[F] - k3[A][60S] \\ \dfrac{d[60S]}{dt} = -k3[A][60S] + k4[R] \\ \dfrac{d[R]}{dt} = k3[A][60S] - k4[R] \\ Psynth(t) = k3[A](t) \end{cases} \quad (5)$$

The model (5) contains three independent conservations laws:

$[F] + [40S] + [A] + [R] = [40S]_0,$

$[F] + [eIF4F] = [eIF4F]_0,$   (6)

$[60S] + [R] = [60S]_0,$

where $[40S]_0$, $[60S]_0$ and $[eIF4F]_0$ are total amounts of available small, big ribosomal subunits and the initiation factor respectively.

The following assumptions on the model parameters were suggested [62]:

$k4 << k1,k2,k3,$

$k3 >> k1,k2,$

$[eIF4F]_0 << [40S]_0,$   (7)

$[eIF4F]_0 < [60S]_0 < [40S]_0,$

with the following justification: ``...The amount 40S ribosomal subunit was set arbitrarily high ... as it is thought to generally not be a limiting factor for translation initiation. In contrast, the level of eIF4F, as the canonical limiting factor, was set significantly lower so translation would be dependent on its concentration as observed experimentally... Finally, the amount of subunit joining factors for the 60S large ribosomal subunit were estimated to be more abundant than eIF4F but still substoichiometric when compared to 40S levels, consistent with in vivo levels... The *k4* rate is relatively slower than the other rates in the model; nevertheless, the simulation's overall protein production was not altered by changes of several orders of magnitude around its value..." [62].

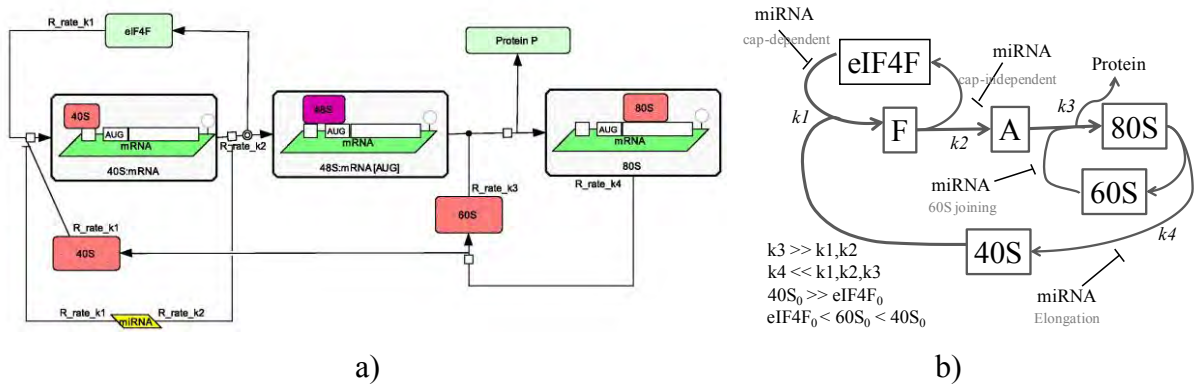

Figure 13.4. The mathematical model of protein translation which explicitly takes into account recycling of ribosomes and initiation factors; a) graphical presentation of the model in the SBGN standard; b) schematic model presentation.

The last statement about the value of *k4* is needed to be made more precise: in the model by Nissan and Parker, *k4* is a sensitive control parameter. It does not affect the steady state protein synthesis rate only in one of the possible scenarios (*inefficient initiation*, deficit of the initiation factors, see below).

The final steady state of the system can be calculated from the conservation laws and the balance equations among all the reaction fluxes:

$$k2[F]_s = k3[A]_s [60S]_s = k4[R]_s = k1[40S]_s [eIF4F]_s \qquad (8)$$

where "*s*" index stands for the steady state value. Let us designate a fraction of the free $[60S]$ ribosomal subunit in the steady state as $x = \frac{[60S]_s}{[60S]_0}$. Then we have

$$[F]_s = \frac{k4}{k2}[60S]_0(1-x), \quad [A]_s = \frac{k4}{k3}\frac{1-x}{x}, \quad [eIF4F]_s = [eIF4F]_s - \frac{k4}{k2}[60S]_0(1-x), \qquad (9)$$

$$[60S]_s = [60S]_0 x, \quad [R]_s = [60S]_0(1-x), \quad [40S]_s = [40S]_0 - [60S]_0(1-x)\left(1+\frac{k4}{k2}\right) - \frac{k4}{k3}\frac{1-x}{x}$$

and the equation to determine *x*, in which we have neglected the terms of smaller order of magnitude, based on conditions (7):

$$x^3 + x^2(\alpha + (\delta-1) + (\beta-1)) + x(-\alpha + (\delta-1)(\beta-1)) + \gamma(1-\beta) = 0, \qquad (10)$$

$$\alpha = \frac{k2}{k1[60S]_0}, \quad \beta = \frac{k2[eIF4F]_0}{k4[60S]_0}, \quad \gamma = \frac{k4}{k3[60S]_0}, \quad \delta = \frac{[40S]_0}{[60S]_0}.$$

From the inequalities on the parameters of the model, we have $\delta > 1$, $\gamma << 1$ and, if $k1 >> k4/[eIF4F]_0$ then $\alpha << \beta$. From these remarks it follows that the constant term $\gamma(1-\beta)$ of the equation (10) should be much smaller than the other polynomial coefficients, and the equation (10) should have one solution close to zero and two others:

$$x_0 \approx \frac{k4}{k3([40S]_0 - [60S]_0)}, \quad x_1 \approx 1 - \frac{k2[eIF4F]_0}{k4[60S]_0} + \frac{k2^2[eIF4F]_0}{k1 \cdot k4 \cdot [40S]_0} \frac{1}{1 - \frac{k2[eIF4F]_0}{k4[40S]_0}}, \quad x_2 \approx 1 - \frac{[40S]_0}{[60S]_0}, \quad (11)$$

provided that α<<|1-δ| or α<<|1-β|. In the expression for $x_1$ we cannot neglect the term proportional to α, to avoid zero values in (10).

The solution $x_2$ is always negative, which means that one can have one positive solution $x_0$<<1, if $\frac{k2[eIF4F]_0}{k4[60S]_0} \geq 1$, and two positive solutions $x_0$ and $x_1$, if $\frac{k2[eIF4F]_0}{k4[60S]_0} \leq 1$. However, it is easy to check that if $x_1 > 0$ then $x_0$ does not correspond to a positive value of $[eIF4F]_s$. This means that for a given combination of parameters satisfying (7) we can have only one steady state (either $x_0$ or $x_1$).

The two values $x = x_0$ and $x = x_1$ correspond to **two different modes of translation**. When, for example, the amount of the initiation factors $[eIF4F]_0$ is **not enough to provide efficient initiation** ($[eIF4F]_0 < \frac{k2}{k4[60S]_0}$, $x = x_1$) then most of the 40S and 60S subunits remain in the free form, the initiation factor eIF4F being always the limiting factor. If the **initiation is efficient** enough ($[eIF4F]_0 > \frac{k2}{k4[60S]_0}$), then we have $x = x_0$<<1 when almost all 60S ribosomal subunits are engaged in the protein elongation, and $[eIF4F]$ being a limiting factor at the early stage. However, it is liberated after and ribosomal subunits recycling becomes limiting in the initiation (see the next section for the analysis of the dynamics).

Let us notice that the steady state protein synthesis rate under these assumptions is

$$Psynth = k4[60S]_0(1-x) = \begin{cases} k4[60S]_0, & \text{if } \frac{k2[eIF4F]_0}{k4[60S]_0} > 1 \\ k2[eIF4F]_0, & \text{else} \end{cases}. \quad (12)$$

This explains the numerical results obtained in [62]: with low concentrations of $[eIF4F]_0$ microRNA action would be efficient only if it affects $k2$ or if it competes with eIF4F for binding to the mRNA cap structure (thus, effectively further reducing the level $[eIF4F]_0$). With higher concentrations of $[eIF4F]_0$, other limiting factors become dominant: $[60S]_0$ (availability of the heavy ribosomal subunit) and $k4$ (speed of ribosomal subunits recycling which is the slowest reaction rate in the system). Interestingly, in any situation the protein translation rate does not depend on the value of $k1$ directly (of course, unless it does not become "globally" rate limiting), but only through competing with eIF4F (which makes the difference with the simplest linear protein translation model).

Equation (12) explains also some experimental results reported in [55]: increasing the concentration of [eIF4F] translation initiation factor enhances protein synthesis but its effect is abruptly saturated above a certain level.

## 5.2. Analysis of the model dynamics

It was proposed to use the following model parameters: $k1=k2=2$, $k3=5$, $k4=1$, $[40S]_0=100$, $[60S]_0=25$, $[eIF4F]_0=6$ [62]. As we have shown in the previous section, there are two scenarios of translation possible in the Nissan and Parker's model which we called "efficient" and "inefficient" initiation. The choice between these two scenarios is determined by the combination of parameters $\beta = \frac{k2[eIF4F]_0}{k4[60S]_0}$. For the original parameters from [62], $\beta = 0.48 < 1$ and this corresponds to the simple one-stage "inefficient" initiation scenario. To illustrate the alternative situation, we changed the value of *k4* parameter, putting it to 0.1, which makes $\beta = 4.8 > 1$. The latter case corresponds to the "efficient" initiation scenario, the dynamics is more complex and goes in three stages (see below).

Simulations of the protein translation model with these parameters and the initial conditions

$$\begin{bmatrix} [40S] \\ [eIF4F] \\ [F] \\ [A] \\ [R] \\ [60S] \end{bmatrix} = \begin{bmatrix} [40S]_0 \\ [eIF4F]_0 \\ 0 \\ 0 \\ 0 \\ [60S]_0 \end{bmatrix}$$

are shown in Figure 13.5. The system shows non-trivial relaxation process which takes place in several epochs. Qualitatively we can distinguish the following stages:

1) Stage 1: Relatively fast relaxation with conditions $[40S]>>[eIF4F]$, $[60S]>>[A]$. During this stage, the two non-linear reactions $40S+eIF4F \rightarrow F$ and $A+60S \rightarrow R$ can be considered as pseudo-monomolecular ones: $eIF4F \rightarrow F$ and $A \rightarrow R$ with rate constants dependent on [40*S*] and [60*S*] respectively. This stage is characterized by rapidly establishing quasiequilibrium of three first reactions (R1, R2 and R3 with *k1*, *k2* and *k3* constants). Biologically, this stage corresponds to the assembling of the translation initiation machinery, scanning for the start codon and assembly of the first full ribosome at the start codon position.

2) Transition between Stage 1 and Stage 2.

3) Stage 2: Relaxation with the conditions $[40S]>> [eIF4F]$, $[60S]<< [A]$. During this stage, the reactions $40S+eIF4F \rightarrow F$ and $A+60S \rightarrow R$ can be considered as pseudo-monomolecular $eIF4F \rightarrow F$ and $60S \rightarrow 80S$. This stage is characterized by two local quasi-steady states established in the two network reaction cycles (formed from R1-R2 and R3-R4 reactions). Biologically, this stage corresponds to the first round of elongation, when first ribosomes move along the coding region of mRNA. The small ribosomal subunit 40*S* is still in excess which keeps the initiation stage (reaction R1-R2 fluxes) relatively fast.

4) Transition between Stage 2 and Stage 3.

5) Stage 3: Relaxation with the conditions $[40S]<< [eIF4F]$, $[60S]<< [A]$. During this stage, the reactions $40S+eIF4F \rightarrow F$ and $A+60S \rightarrow R$ can be considered as pseudo-monomolecular $40S \rightarrow F$ and $60S \rightarrow R$. During this stage all reaction fluxes are balanced. Biologically, this stage corresponds to the stable production of the protein with constant recycling of the ribosomal subunits. Most of ribosomal subunits 40*S* are involved in protein elongation, so the initiation process should wait the end of elongation for that they would be recycled.

Our analysis of the non-linear Nissan and Parker's model showed that the protein translation machinery can function in two qualitatively different modes, determined by the ratio $\beta = \frac{k2[eIF4F]_0}{k4[60S]_0}$ [90]. We call these two modes "efficient initiation" ($\beta>1$) and "inefficient initiation" ($\beta<1$) scenarios. Very roughly, this ratio determines the balance between the overall speeds of initiation and elongation processes. In the case of "efficient initiation" the rate of protein synthesis is limited by the speed of recycling of the ribosomal components

(60S). In the case of "inefficient initiation" the rate of protein synthesis is limited by the speed of recycling of the initiation factors (eIF4F). Switching between two modes of translation can be achieved by changing the availability of the corresponding molecules ( $[60S]_0$ or $[eIF4F]_0$ ) or by changing the sensitive kinetic parameters (*k2* or *k4*).

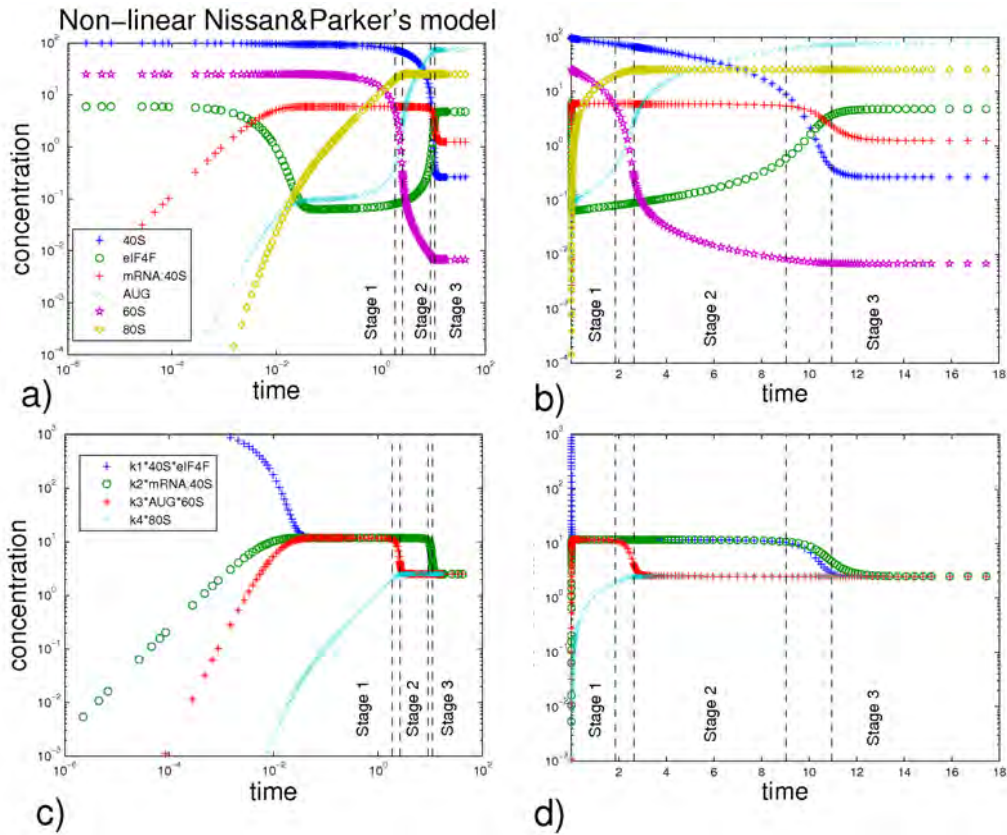

Figure 13.5. Numerical simulations of the species concentrations and fluxes of the non-linear translation model. a) and c) log-log scale; b) and d) log-scale in values, and linear scale for the time axis (the time units are arbitrary, since the dimensionality of the parameters in [62] was not specified).

As a result of the dynamical analysis, we assembled an approximate solution of the non-linear system under assumptions (7) about the parameters. The detailed description of this solution is given in [90]. The advantage of such a semi-analytical solution is that one can predict the effect of changing the system parameters.

One of the obvious predictions is that the dynamics of the system is not sensitive to variations of *k3*, so if microRNA acts on the translation stage controlled by *k3* then no microRNA effect could be observed looking at the system dynamics (being the fastest one, *k3* is not a sensitive parameter in any scenario).

If microRNA acts on the translation stage controlled by *k4* (for example, by ribosome stalling mechanism) then we should consider two cases of efficient (β>1) and inefficient (β <1) initiation. In the first case the steady state protein synthesis rate is controlled by *k4* (as the slowest, limiting step) and any effect on *k4* would lead to the proportional change in the steady state of protein production. By contrast, in the case of inefficient initiation, the steady state protein synthesis is not affected by *k4*. Instead, the relaxation time is affected, being $\sim \frac{1}{k4}$. However, diminishing *k4* increases the β parameter: hence, this changes

"inefficient initiation" scenario for the opposite, making *k4* sensitive for the steady state protein synthesis anyway, when *k4* becomes smaller than $\frac{k2[eIF4F]_0}{[60S]_0}$. For example, for the default parameters of the model, decreasing *k4* value firstly leads to no change in the steady state rate of protein synthesis, whereas the relaxation time increases and, secondly, after the threshold value $\frac{k2[eIF4F]_0}{[60S]_0}$ starts to affect the steady state protein synthesis rate directly. This is in contradiction to the message from [62] that the change in *k4* by several orders of magnitude does not change the steady state rate of protein synthesis.

Analogously, decreasing the value of *k2* can convert the "efficient" initiation scenario into the opposite after the threshold value $\frac{k4[60S]_0}{[eIF4F]_0}$. We can recapitulate the effect of decreasing *k2* in the following way: 1) in the case of the "efficient" initiation *k2* does not affect the steady state protein synthesis rate up to the threshold value after which it affects it in a linear manner. The relaxation time drastically increases, because decreasing *k2* leads to elongation of all dynamical stages duration (for example, we have estimated the time of the end of the dynamical Stage 2 as $t''' = \frac{[40S]_0}{k2[eIF4F]_0}$. However, after the threshold value the relaxation time decreases together with *k2*, quickly dropping to its unperturbed value; 2) in the case of "inefficient" initiation the steady state protein synthesis rate depends proportionally on the value of *k2* (12), while the relaxation time is not affected.

Table 13.2. Modeling of four mechanisms of microRNA action in the non-linear protein translation model

| Observable value | Initiation | Step after initiation | Ribosome assembly | Elongation |
|---|---|---|---|---|
| **Wild-type cap, inefficient initiation** | | | | |
| *Steady-state rate* | slightly decreases | decreases | no change | decreases after threshold |
| *Relaxation time* | no change | no change | no change | goes up and down |
| **Wild-type cap, efficient initiation** | | | | |
| *Steady-state rate* | no change | slightly decreases after strong inhibition | no change | decreases |
| *Relaxation time* | no change | goes up and down | no change | no change |
| **A-cap, inefficient initiation** | | | | |
| *Steady-state rate* | decreases | decreases | no change | slightly decreases after strong inhibition |
| *Relaxation time* | no change | no change | no change | goes up and down |
| **A-cap, efficient initiation** | | | | |
| *Steady-state rate* | decreases after threshold | slightly decreases after strong inhibition | no change | decreases |
| *Relaxation time* | goes up and down | goes up and down | no change | increases |

MicroRNA action on *k1* directly does not produce any strong effect neither on the relaxation time nor on the steady state protein synthesis rate. This is why in the original work [62] cap-dependent mechanism of microRNA action was taken into account through effective change of the $[eIF4F]_0$ value (total concentration of the translation initiation factors), which is a sensitive parameter of the model (5).

The effect of microRNA through various mechanisms and in various experimental settings (excess or deficit of eIF4F, normal cap or A-cap) is recapitulated in Table 13.2. The conclusion that can be made from this table is that all four mechanisms show clearly different patterns of behavior in various experimental settings. From the simulations one can make a conclusion that it is still not possible to distinguish between the situation when microRNA does not have any effect on protein translation and the situation when it acts on the step which is neither rate limiting nor "second rate limiting" in any experimental setting ($k3$ in our case). Nevertheless, if any change in the steady-state protein synthesis or the relaxation time is observed, theoretically, it is possible to specify the mechanism responsible for it.

## 6. General model of miRNA-mediated translation regulation

Nine distinct mechanisms of microRNA action have been described in the literature: the main experimental data supporting each proposed mechanism are summarized in the review section of this chapter. The complete model containing all known microRNA action mechanisms is shown in Figure 13.6a using an SBGN standard diagram.

The principal differences between the Nissan and Parker's model and the model described in this section are 1) the complete model describes all nine known mechanisms of miRNA action; 2) mRNA amount is a dynamical variable, i.e. it is modelled explicitly, taking into account its synthesis and degradation; 3) we explicitly model binding of miRNA at various stages of translation, i.e. in our model both mRNA in free and miRNA-bound forms are present; 4) we assume concentration of eIF4F and ribosomal subunits present in excess, as in the simplest model.

For modelling, we assumed that the initiation factors and ribosomal subunits are always available in excess. This allowed us to simplify the model to 12 chemical species and 20 reactions, as described below and schematically shown in Figure 13.6b:

[M0] – new synthesized and not yet initiated mRNA

[F0] – new initiated mRNA, with initiation complex, including 40S ribosomal subunit

[M] – initiated mRNA with free translation initiation site

[F] – initiated mRNA with translation initiation site occupied by 40S ribosomal subunit

[R] – number of ribosomes fully assembled on miRNA-free mRNA

[M'0] – new synthesized not initiated mRNA with one or more miRNAs bound

[F'0] – new mRNA with initiation complex, including 40S ribosomal subunit, with miRNA(s) bound to mRNA

[M'] – initiated miRNA-bound mRNA with free translation initiation site

[F'] – initiated miRNA-bound mRNA with translation initiation site occupied by 40S ribosomal subunit

[R'] – ribosomes fully assembled on miRNA-bound mRNA

[P] – protein, completely translated from the given mRNA

[B] – mRNA sequestered in P bodies.

Let us make a notice on interpretation of some of the model variables. Explicit description of mRNA:ribosome complexes would require separate dynamical variables for the amounts of mRNA with one ribosome, mRNA with two ribosomes, mRNA with three ribosomes, and so on (potentially, large number of variables). To avoid this complexity, we apply lumping of the detailed model, described in [24]. In the lumped reaction network, new produced mRNA (state M0) is first initiated and prepared for the first round of translation (state F0). After that, the initiated mRNA alternates between states M (state ready for the next round of translation) and F (mRNA prepared for the next ribosome assembly). During each such a round, a new assembled ribosomal complex (R) appears in the system. Thus, we explicitly separate the process of mRNA initiation (which can include capping, adenylylation, circularization, mRNA transport to specific cellular regions) and the process of recruiting 40S ribosomal subunit at already initiated mRNA for the next round of translation. In our model, these two processes proceed with different speeds.

In our interpretation, we consider mRNAs as places for a catalytic reaction (protein translation). These places (amount of catalyzer) in our model can be synthesized or destroyed and present in four states (non-initiated, initiated, in 'translating' state ready for assembling new ribosome and in 'translating' state with a new assembling ribosome). To take into account miRNA, we say that there are two types of catalyzer: miRNA-free and miRNA-bounded, with different rate constants of degradation. miRNA-free catalyzer can be irreversibly transformed into miRNA-bounded type of catalyzer.

Importantly, [R] in our interpretation is not the amount of mRNA translating proteins but the amount of ribosomes bound to mRNA and translating proteins, i.e. the number of sites where the catalysis takes place. Dividing the number of these sites on the amount of the catalyzer in the initiated state [M]+[F] gives the average number of ribosomes per translating mRNA, which we denote as [RB].

The definition of the kinetic rate constants used further in the paper is the following:

1. null → $M_0$, the free mRNA is transcribed in the system with the rate constant $k_t$.

2. $M_0$ → $F_0$, assembly of initiation complex and 40S ribosomal subunit with mRNA occurs with the rate constant $k_{01}$

3. $F_0$ → M+R, assembly of the first ribosome on the initiation site with the rate constant $k_2$

4. M → F, initiation of the translation (recruitment of 40S subunit) on already translated mRNA, with the rate constant $k_1$

5. F → M+R, assembly of full ribosome (S80) on mRNA occurs with the rate constant $k_2$

6. R → P, translation of the protein with consequent release of ribosomes occurs with the rate constant $k_3$

7. R → null, degradation of mRNA leads to release of ribosomes with the rate constant $k_d$, same reaction describes premature ribosome drop-off from mRNA with the rate constant $k_{rd}$

We will assume that the process of microRNA binding to mRNA can occur at various stages of translation and that its rate $k_b$ will be the same in each of the following reactions:

8. $M_0$ → $M_0'$

9. $F_0$ → $F_0'$

10. M → M'

11. F → F'

12. R → R'

In the same way we will assume that the rate of degradation of mRNA not driven by microRNA action ($k_d$) can be considered as the same one at all stages of translation:

13. $M_0$ → null

14. $F_0$ → null

13. M → null

14. F → null

The degradation rate of mRNA bound to microRNA could occur with or without direct action of microRNA on its degradation. For the beginning we will assume that this rate constant ($k_d'$) is different from the free mRNA degradation and it is the same one for all stages of translation:

15. $M'_0$ → null

16. $F'_0$ → null

17. M' → null

18. F' → null

19. R' → null

Next we assume that the reaction corresponding to the assembly of the initiation complex and 40S ribosomal subunit with mRNA in the presence of miRNA ($M'_0$ → $F'_0$) will occur with the rate constant $k_{01}'$.

20. $M'_0$ → $F'_0$

Recruitment of 40S subunit on already translating miRNA-bound mRNA occurs with the rate constant $k_1'$:

21. M' → F'

Reactions of assembly of the full ribosome (S80) on mRNA in the presence of microRNA occur with the rate constant $k_2'$:

22. $F'_0$ → M'+R

23. F' → M'+R

The rate of protein production in the case of microRNA action is described by the following reaction:

24. R' → P, with the rate constant $k_3'$

25. R' → null, describes possible mechanism of ribosomal drop-off (without protein production), with the rate constant $k_{rd}'$, and mRNA degradation with ribosome release, with the rate constant $k_d'$.

Reactions 26 and 27 describe the reverse process of mRNA sequestration in P-bodies, with rates $k_{+s}$ and $k_{-s}$ correspondingly:

26. $M_0'$ → B,

27. B → $M_0'$

The mRNA in P-bodies is degraded with specific rate $k_{bd}'$

28. B → null

29. P + R' → null, the rate of protein degradation by microRNA-independent mechanisms is $k_p$, while it can be increased in the presence of miRNA by $k_r \times R'$.

The system of equations $dx/dt = \mathbf{K}_0 + \mathbf{K}x$ (where $x$ is the vector of 12 dynamic variables, $\mathbf{K}$ is the kinetic matrix, and $\mathbf{K}_0$ is the vector of production with only one non-zero component corresponding to the transcriptional synthesis of mRNA) has the following form:

Thus, for this simplified linear model we need to define the proper values for 18 coefficients corresponding to the rates of reactions.

For simulations, we needed the numerical values of 18 kinetic coefficients, which were estimated from published reports and are provided in Table 13.3. Although it is obvious that all rates diverge considerably for different mRNAs, experimental data mining allowed us to make a plausible assumption for almost all of the kinetic rates used in the model. For example, mRNA half-lives vary from a few minutes to more than 24 h, with a mean at 10 h [88], which we selected as the corresponding rate. It is nevertheless possible that highly regulated mRNAs, such as most miRNA targets, have shorter half-lives. The same reasoning also applies to protein half-lives.

$$\begin{cases} \dfrac{d[M_0]}{dt} = k_t - (k_d + k_{01} + k_b)[M_0] \\ \dfrac{d[F_0]}{dt} = k_{01}[M_0] - (k_d + k_2 + k_b)[F_0] \\ \dfrac{d[M]}{dt} = k_2([F_0] + [F]) - (k_d + k_1 + k_b)[M] \\ \dfrac{d[F]}{dt} = k_1[M] - (k_d + k_2 + k_b)[F] \\ \dfrac{d[R]}{dt} = k_2([F_0] + [F]) - (k_d + k_{rd} + k_3 + k_b)[R] \\ \dfrac{d[M'_0]}{dt} = k_b[M_0] - (k'_d + k'_{01})[M'_0] - (k_{+s}[M] - k_{-s}[B]) \\ \dfrac{d[F'_0]}{dt} = k_b[F_0] + k'_{01}[M'_0] - (k'_d + k'_2)[F'_0] \\ \dfrac{d[M']}{dt} = k_b[M] + k'_2([F'_0] + [F']) - (k'_d + k'_1)[M'] \\ \dfrac{d[F']}{dt} = k_b[F] + k'_1[M] - (k'_d + k'_2)[F'] \\ \dfrac{d[R']}{dt} = k_b[R] + k'_2([F_0] + [F]) - (k'_d + k'_{rd} + k'_3)[R] \\ \dfrac{d[P]}{dt} = k_3[R] + k'_3[R'] - (k_p + k_r[R'])[P] \\ \dfrac{d[B]}{dt} = k_{+s}[M'] - k_{-s}[B] - k_{bd}[B] \end{cases}$$

Similarly, we estimated the elongation time for mRNA translation as 1-2 min [7,30,76], even though it depends on the mRNA length: at 10 aa/sec [20], 1-2 min corresponds to a mean length of 1.8 to 3.6 kb [26]. Likewise, the numbers of ribosomes per mRNA molecule are highly variable, from 4-5 to more than 10 [7,54]. We considered 6 ribosomes per mRNA as being a reasonable assumption. We therefore postulated that 6 initiation events occur during a cycle of elongation, which leads to an estimate of 6 initiations/minute, and is of the same order of magnitude as what has been proposed previously [7]. All information concerning the kinetic coefficients we used for our modelling is summarized in Table 13.3.

# 7. Distinct dynamical types of miRNA action and kinetic signatures of miRNA mechanisms

## 7.1. Analytical solution of model equations for the case of normal translation (no miRNA)

The dynamical variables that can be observed and measured in the experiment are

Total amount of mRNA:   **MT** = $M_0$+$F_0$+M+F+$M_0$'+$F_0$'+M'+F'+B

Total amount of protein:   **PR** = P

Average number of ribosomes,
translating one mRNA:   **RB** = (R+R')/( M+M'+F+F')

The solution of model equations and expression for [MT], [PR], [RB] for the trivial case without miRNA in the system were obtained. This can be modelled by putting to zero the binding constant $k_b = 0$.

For this case the steady state values for the measurable quantities are

$$MT^{SS} = \frac{k_t}{k_d}, \quad RB^{SS} = \frac{k_2+k_d}{k_3+k_d+k_{rd}}, \quad PT^{SS} = \frac{k_3}{k_p}\frac{k_t}{k_d}\frac{k_{01}k_2}{(k_{01}+k_d)(k_3+k_d)} \quad (14)$$

and the relaxation times are

$$MT^{RT} = \frac{1}{k_d}, \quad RB^{RT} = \frac{1}{\min(k_{01}+k_d, k_2+k_d, k_3+k_d+k_{rd})}, \quad PT^{RT} = \frac{1}{\min(k_d, k_p)}, \quad (15)$$

where we have assumed that $k_1 \gg k_{01}, k_2, k_3$.

These formulas allow qualitative understanding of the effect of miRNA on various steps of translation and the corresponding kinetic signatures. They can also help to decipher experimentally observed kinetic signatures when multiple mechanisms are present simultaneously and the translation parameters are not known. Exact recipe on doing this will be a subject of our future work.

## 7.2. Dominant paths of the model and their relations to the miRNA mechanisms

According to the methodology of asymptotology [23], let us consider the case of well separated constants, i.e. when any two kinetic constants in the graph in the Figure 13.6b have different orders of magnitude at each fork (i.e., a node with several outgoing reactions). Each such a (partial) ordering of kinetic constants will generate a path on the graph (possibly, cyclic), starting at $M_0$ node. We will call it **the dominant path**. Each path corresponds to one (if it does not contain cycles) or several (if it contains a cycle) dominant systems and to a distinguishable biochemical scenario. For example, the partial ordering ($k_b \gg k_1, k_d$ ; $k'_{01} \gg k_{-s}, k'_d$ ; $k'_2 \ll k'_d$) corresponds to the dominant path describing the process of translation inhibition via 60S subunit joining repression (see Table 13.4, path $M_0M'_0F'_0$).

A dominant path is connected to a dominant system (whose solution of the corresponding dynamics equations provides an asymptotic approximation of the whole system dynamics) in the following way. If the path does not contain cycles, then it represents the dominant system. If the path contains cycles then the cycles should be glued and represented by single nodes (which will represent quasistationary distribution of chemical species concentrations inside the cycle). Then one should find the dominant path for the new graph with glued cycles and continue until an acyclic dominant path will be found. Depending on the ordering of kinetic rates inside each cycle, one cyclic dominant path can lead to several different dominant systems. The dominant system in general represents a hierarchy of glued cycles. The details of constructing dominant systems are provided in [22,72].

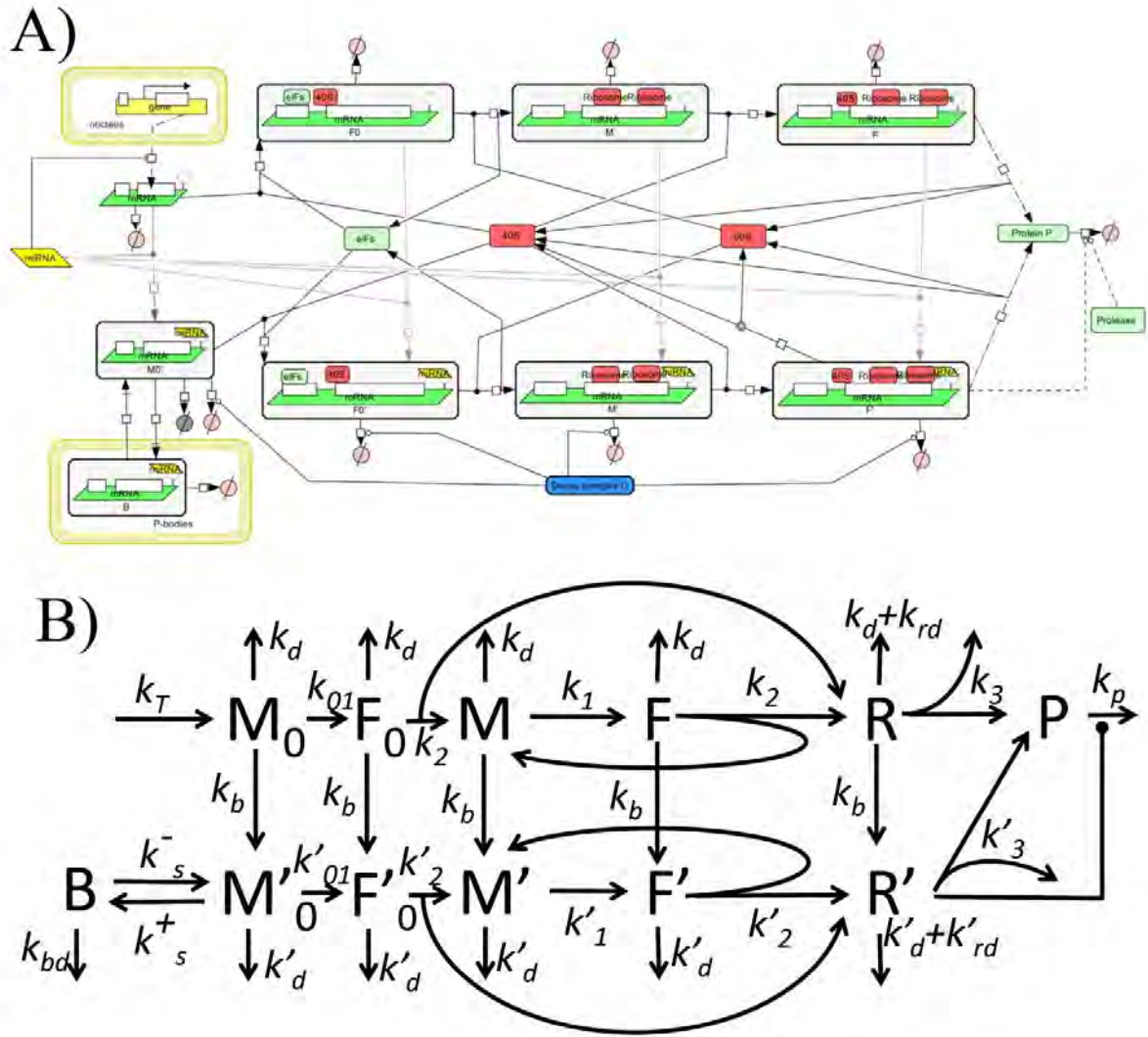

Figure 13.6. Mathematical model taking into account all nine mechanisms of miRNA action; a) graphical presentation of the model in the SBGN standard; b) schematic model presentation in the assumption that ribosomal subunits and initiation factors are present in excess.

It is convenient to designate each dominant path by nodes through which it passes. There are many possible reaction graph traversals leading to multiple possible dominant paths, if one considers all partial orderings of the constants in the reaction forks. However, some of them are biologically non-relevant. For example, the ordering $k_d \gg k_{01}$ (dominant path $M_0$) will not lead to any translation (the mRNA will be degraded before it will be initiated). In the same way, $k_d \gg k_2$ (dominant path $M_0F_0$) will terminate the normal translation prematurely. Thus, we postulate $k_d \ll k_{01}, k_2, k_3$. Also for simplicity we assume that binding of miRNA to mRNA is more rapid than normal initiation, i.e., $k_b \gg k_{01}, k_2, k_3$ if there is miRNA in the system, and $k_b = 0$, if not. Also we assume that $k_{01} \ll k_1$, because $k_1$ corresponds to recruiting 40S subunit on already initiated and translated mRNA (which we assume never be rate-limiting), while $k_{01}$ includes both mRNA initiation and 40S subunit recruiting. This leads to 6 biologically relevant dominant paths, all of which are listed in Table 13.4.

Table 13.3. Reference set of parameters for the model and their changes according to the action of various miRNA-mediated mechanisms of translation repression

| Kinetic rate constant | Reference value or interval | Comment |
|---|---|---|
| **Parameters of transcription and translation without miRNA action** | | |
| $k_t$ | $10^{-3}$ | *Transcription kinetic rate.* If **Transcriptional Inhibition** mechanism is active then this constant is proportionally reduced from $k_t$ (0% efficiency of the mechanism) to zero (100% efficiency of the mechanism). |
| $k_{01}$ | $2 \cdot 10^{-4}$ | *mRNA early initiation rate in the absence of miRNA.* |
| $k_1$ | 1 | *rate of 40S recruitement at already translated mRNA, considered to be fast and not rate-limiting* |
| $k_2$ | $6 \cdot 10^{-2}$ | *60S unit joining and assembly of the full ribosome on mRNA rate in the absence of miRNA.* |
| $k_3$ | $10^{-2}$ | *Rate including elongation and termination of translation in the absence of miRNA.* In all simulations of translation without miRNA, we assume that $k_3 = k_3/6$, which gives 6 ribosomes sitting on one translated mRNA in average. |
| $k_d$ | $10^{-5}$ | *mRNA degradation rate in the absence of miRNA.* In all simulations of translation without miRNA, we assume that $k_d << k_1, k_2, k_3$. Otherwise mRNA will be degraded much faster than it will be initiated and translated. |
| $k_{rd}$ | 0 | *Rate of ribosome drop-off.* We neglect the ribosome drop-off in the absence of miRNA |
| $k_p$ | $5 \cdot 10^{-6}$ | *Rate of protein degradation in the absence of miRNA.* |
| **Parameters of various mechanisms of miRNA action** | | |
| $k_b$ | $10^{-3}$ (strong) $10^{-4}$ (medium) $10^{-5}$ (weak) | *Rate of miRNA binding to mRNA.* This rate depends on many factors including the complementarity of miRNA sequence to the sequence of the binding site. We assume that depending on these factors, the rate can vary in the range of several orders of magnitude. When $k_b <<\min(k_1, k_2, k_3)$, we consider the binding as weak, because it does not considerably influence the rate of translation. |
| $k_{01}'$ | $[0; k_{01}]$ | *mRNA initiation rate with miRNA.* If **Cap Inhibition** mechanism is active then this constant can be proportionally reduced from $k_1$ to zero. |
| $k_1'$ | $k_{01}$ | *40S recruitement at already translated miRNA-bound mRNA,* we do not consider the corresponding hypothetical mechanism in the model |
| $k_2'$ | $[0; k_2]$ | *60S unit joining and assembly of the full ribosome on mRNA rate with miRNA.* If **60S Unit Joining Inhibition** mechanism is active then this constant can be proportionally reduced from $k_2$ to zero. |
| $k_3'$ | $[0; k_3]$ | *Rate including elongation and termination of translation with miRNA.* If **Elongation Inhibition** mechanism is active then this constant can be proportionally reduced from $k_3$ to zero. |
| $k_d'$ | $[k_d; 10^2 \cdot k_d]$ | *Rate of mRNA degradation with miRNA.* If **Decay** mechanism is active then this constant can increase 10-fold at 100% mechanism efficiency. If **Cleavage** mechanism is active then this constant can increase by 100-fold. |
| $k_{\pm s}$ | $[0; 5 \cdot 10^{-2}]$ | *Rate of reversible capturing of mRNA to P-bodies.* If **P-bodies Sequestration** mechanism is active, this constant can be proportionally increased from zero to $k_{+s}$. The reverse rate constant $k_{-s}$ is assumed to be $k_{-s} = 5 \cdot k_{+s}$. We assume that mRNA can be degraded in P-bodies with the rate $k_d'$. |
| $k_{rd}'$ | $[0; 5 \cdot k_3']$ | *Rate of ribosome drop-off.* If **Ribosome Drop-Off** mechanism is active then this constant is proportionally increased from 0 to $5 \cdot k_3'$. |
| $k_r$ | $[0; 5 \cdot 10^{-5}]$ | *Rate of co-translational protein degradation catalysis.* If **Co-Translational Protein Degradation** mechanism is active then this constant is proportionally increased from zero to $5 \cdot 10^{-5}$, and the protein degradation rate is increased as $k_p^{miRNA} = k_p + k_r \cdot \mathbf{R'}$. |

Table 13.4 shows that the types of dynamical behavior (dominant paths) can be mapped onto the biologically characterized mechanisms of miRNA action, but this mapping is not one-to-one: several biological mechanisms can correspond to one dynamical type (for example, $M_0M'_0$ dominant path corresponds to M1, M7 and M8 biological mechanisms and, conversely, biological mechanism M1 can correspond to $M_0M'_0$ or $M_0M'_0F'_0M'F'R'P$ dominant paths).

### 7.3. Kinetic signatures of miRNA-mediated mechanisms of protein translation inhibition

In order to provide a practical recipe to distinguish between nine different mechanisms of miRNA action, we studied the dynamical behaviour of the model for the reference set of parameters for weak, medium and strong miRNA binding strengths. The simulation was performed in the following way:

1) First, the system was simulated from zero initial conditions without presence of miRNA ($k_b = 0$) in the time interval $[0; 20/k_d]$. The steady state and relaxation time values for MT, RB and PR values were estimated from the simulation.

2) The miRNA binding constant was changed to the corresponding value and the simulation was continued from the steady state obtained before in the time interval $[20/k_d; 40/k_d]$. New steady state and relaxation time values were estimated from the simulation.

The model includes a vector of parameters $P=\{k_t, k_{01}, k_1, k_2, k_3, k_d, k_p\}$ and of mechanism strength spectrum $S=\{s_1,s_2,\ldots,s_9\}$ (see the next section), which can vary. Each computational experiment is defined by the corresponding vectors $P$ and $S$, binding constant for miRNA ($k_b$) the rest of the model parameters is computed using the following formulas:

| | |
|---|---|
| M1 (Cap Inhibition): | $k_{01}' := (1-s_1) \cdot k_{01}$, |
| M2 (60S Unit Joining Inhibition): | $k_2' := (1-s_2) \cdot k_2$, |
| M3 (Elongation Inhibition): | $k_3' := (1-s_3) \cdot k_3$, |
| M4 (Ribosome Drop-Off): | $k_{rd} := 5 \cdot s_4 \cdot k_3'$, |
| M5 (Co-translational protein degradation): | $k_r := s_5 \cdot k_r^{(ref)}$. |
| M6 (Sequestration in P-bodies): | $k_{+s} := 5 \cdot s_6 \cdot k_s^{(ref)}$, $k_{-s} = s_6 \cdot k_s^{(ref)}$, |
| M7 (Decay of mRNA): | $k_d' := (1+9 \cdot s_7) \cdot k_d$, |
| M8 (Cleavage of mRNA): | $k_d' := (1+99 \cdot s_8) \cdot k_d$, $k_b' := (1+99 \cdot s_8) \cdot k_b$ |
| M9 (Transcriptional Inhibition): | $k_t' := (1-s_9) \cdot k_t$, |

The result of the simulation is a kinetic signature for a mixed mechanism of miRNA action, characterized by six numbers: relative changes of the steady states $MT^{SS} = \frac{MT^{SS}_{miRNA}}{MT^{SS}_{no\ miRNA}}$,

$RB^{SS} = \frac{RB^{SS}_{miRNA}}{RB^{SS}_{no\ miRNA}}$, $PR^{SS} = \frac{PR^{SS}_{miRNA}}{PR^{SS}_{no\ miRNA}}$ and relative changes of relaxation times $MT^{RT} = \frac{MT^{RT}_{miRNA}}{MT^{RT}_{no\ miRNA}}$, $RB^{RT} = \frac{RB^{RT}_{miRNA}}{RB^{RT}_{no\ miRNA}}$, $PR^{RT} = \frac{PR^{RT}_{miRNA}}{PR^{RT}_{no\ miRNA}}$. For the further analysis, using Principal Component Analysis (PCA) we use the logarithms of these ratios.

First, we considered only "pure" mechanisms acting at the maximum 100% efficiency (which leads, for example, for a complete block of mRNA elongation in the presence of miRNA, for

**Elongation Inhibition** mechanism). The resulting signatures are shown in Figure 13.7. Several conclusions can be made from it.

Firstly, the signatures of nine mechanisms are *qualitatively different*, i.e. they can be reliably distinguished in principle, if the 6 required numbers would be estimated experimentally.

Secondly, not all mechanisms can be distinguished only based on the steady-state value analysis, in accordance with the results of modelling described in the previous sections. Some of the relaxation time relative changes should be measured as well in order to distinguish, for example, Ribosome Drop-Off from 60S Unit Joining Inhibition.

Thirdly, one can observe that some of the signature components strongly depend in the quantitative fashion on the order of the miRNA binding constant, and some are completely insensitive. This suggest an experiment in which several sequences of miRNA would be utilised having different (weak, medium, tight) affinities to the target mRNA binding site. Observing how the dynamics of observable quantities are changing with the binding affinity, one can distinguish the mechanisms more reliably. For example, in the case of Ribosome Drop-Off the ribosomal profile should be more sensitive to changing miRNA affinity compared to 60S Unit Joining.

## 8. Coexistence of multiple mechanisms of miRNA action

One of the most debated questions on the action of miRNA on translation is the possibility of co-existence of several mechanisms of miRNA action. Let us study formally to what consequences it can lead from the point of view of translation dynamics and kinetic signatures.

We formalize co-existence of several miRNA action mechanisms in the following way. We characterized a situation when a miRNA can interfere with several steps of translation (and transcription) by a *strength spectrum* of 9 "pure" mechanisms. The spectrum is a 9-dimensional vector $S=\{s_1,s_2,\ldots,s_9\}$ with components corresponding to the strengths (contributions) of "pure" mechanisms M1, M2,…, M9. Each strength $s_i$ of this vector can vary from 0.0 (absence of the mechanism) to 1.0 (or 100%, maximum strength of the mechanism). We call this situation a "combined" mechanism of miRNA action. In this sense, the "pure" mechanisms acting at maximum strength (1.0) are basis vectors in the space of "combined" mechanisms. For example, the spectrum $S=\{0,1.0,0,0,0,0,0,0,0\}$ corresponds to the blockage of 60S unit joining by miRNA without affecting any other step of translation, while $S=\{0.8,0,0,0.5,0,0,0,0,0\}$ corresponds to co-existence of Cap Inhibition (at 80% of its maximal strength) and Ribosome drop-off (at 50% of its maximal strength). Also there are 7 normal translation parameters (without miRNA) $k_t, k_{01}, k_1, k_2, k_3, k_d, k_p$, which allow to consider a vector of parameters $P=\{k_t, k_{01}, k_1, k_2, k_3, k_d, k_p\}$ in 7–dimensional space of parameters.

In this section we make two computational experiments in which we exhaustively study the effect of 1) varying *S* given *P* fixed at reference parameters; and 2) varying *P* given *S*, for four mostly referenced mechanisms: Cap Inhibition, 60S Unit Joining Inhibition, Elongation Inhibition, Decay. In other words, in the first case we study the effect of co-existence of various mechanisms for a given experimental system, characterized by a given set of normal translation parameters. In the second case, we study the effect of variable experimental (or cellular) conditions on the conclusions one can make for the same mixed mechanism of miRNA action. Thus, the results of this section generalize the results of the previous sections to the case of co-existence of several mechanisms at the same time.

Table 13.4. Dominant paths of the unified model of microRNA action mechanisms

| Dominant path | Biological interpretation | Corresponding miRNA-mediated translation repression mechanism(s) |
|---|---|---|
| $M_0F_0MFRP$ | $M_0F_0MFRP$ normal translation with negligible effect of miRNA | None |
| $M_0M'_0$ | $M_0M'_0$ the dominant effect is degradation of mRNA by miRNA | M1: Cap inhibition<br>M7: Decay<br>M8: Cleavage |
| $M_0M'_0B$ | $M_0M'_0B$ mRNA is captured in P-bodies | M6: Sequestration of mRNA in P-Bodies |
| $M_0M'_0F'_0$ | $M_0M'_0F'_0$ mRNA translation is stuck after initiation, before the assembly of the ribosome | M2: 60S subunit joining inhibition |
| $M_0M'_0F'_0M'F'R'$ | $M_0M'_0F'_0M'F'R'$ mRNA is stuck with ribosomes on it and destroyed, or mRNA translation is prematurely aborted | M3: Elongation inhibition<br>M4: Ribosome drop-off |
| $M_0M'_0F'_0M'F'R'P$ | $M_0M'_0F'_0M'F'R'P$ protein synthesis in the presence of miRNA with low mRNA degradation | M1: Cap inhibition<br>M2: 60S subunit joining inhibition<br>M3: Elongation inhibition<br>M5: Co-translational protein degradation mechanisms |

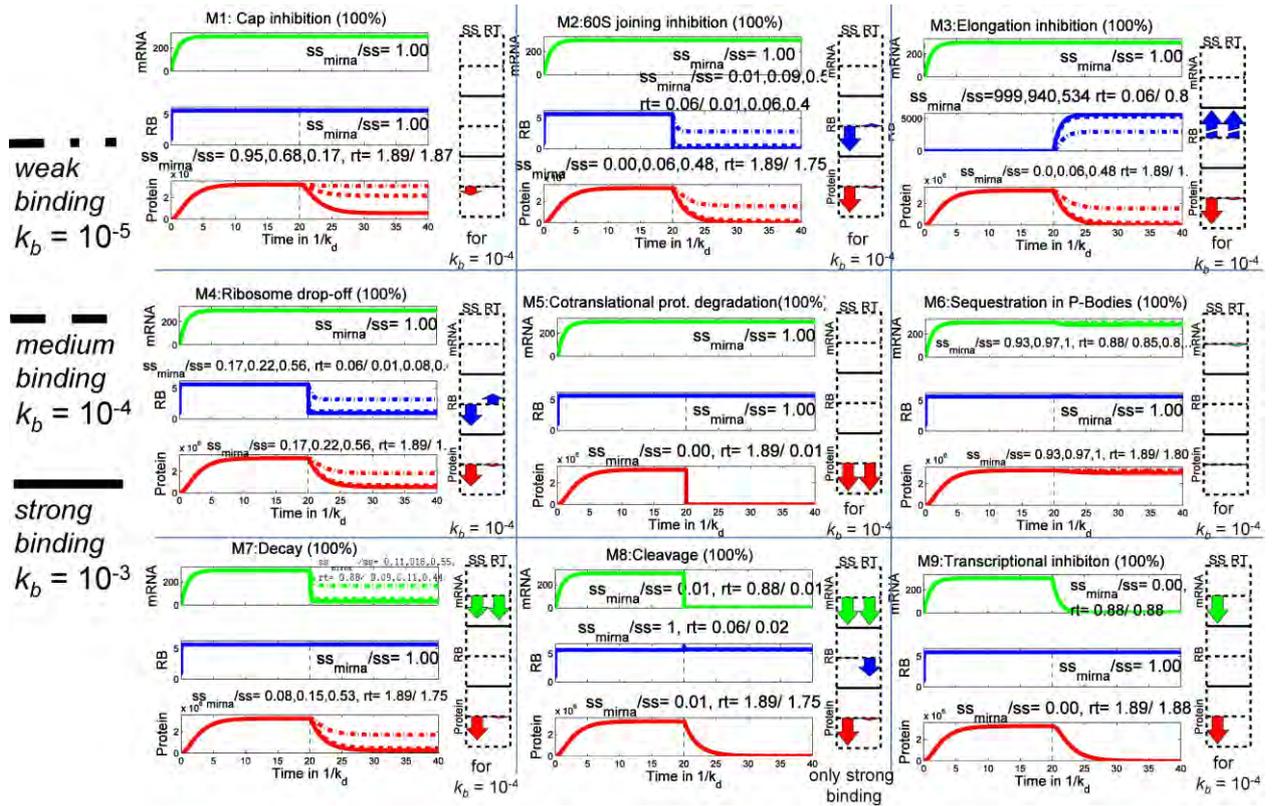

Figure 13.7. Kinetic signatures of the mechanisms of miRNA action. There are nine signatures corresponding to nine mechanisms. Each plot shows dynamics of three quantities: amount of mRNA (*mRNA*), average number of ribosomes per translated mRNA (*RB*), total amount of protein (*Protein*) in the time units measured in $1/k_d$. The dynamics on the left from the dashed line shows translation without miRNA which is added at the time point 20. Three scenarios are simulated for each signature: strong, medium and weak binding strength of miRNA to mRNA. The numbers on the graphs shows relative change in the steady state ($ss_{miRNA}/ss$) and change in the relaxation time (*rt*, measured in $1/k_d$). If three numbers are shown separated by comma, they correspond to weak, medium and strong miRNA binding. If only one number is shown, it means that the binding strength does not affect this quantity significantly. The diagrams on the right from the dynamics plot visualize values of six numbers (relative changes of steady state (SS) and relaxation time (RT) for three measurable quantities) for the case of medium binding strength.

### 8.1. Fixed set of translation parameters and variable mixed mechanisms of miRNA action

In Figure 13.8 we present the results of the following computational experiment. For a reference set of parameters (Table 13.3) we computed 625 kinetic signatures corresponding to all possible combinations of four mechanism strengths ($s_1$, $s_2$, $s_3$, $s_7$) at the level of 0%, 25%, 50%, 75% and 100%. The signatures can be represented as a cloud of 625 points in the 6-dimensional space of kinetic signatures, which was projected on a 2D plane using the standard principal components analysis (PCA). From the Figure 13.8 one can conclude that the first principal component PC1 is mainly associated with the change of ribosomal profile, while the second is mainly associated with degradation of mRNA. Therefore, position of "pure" mechanisms 60S Unit Joining Inhibition and Elongation Inhibition is placed at the maximum distance on the plot, while Cap Inhibition and Decay is located quite closely, because both do not change the ribosomal profile. However, one can show that Cap Inhibition and Decay pure mechanisms are separated along the third principle direction PC3, invisible on the plot.

One of the important conclusions that can be made from the plot in the Figure 13.8 is that the presence of Decay mechanism in the spectrum ($s_7 > 0$) can mask the effect of other mechanisms leading to the very early blockage of translation ($M_0M_0'$ dominant path). Indeed, it might not matter that a translation in the presence of miRNA is completely blocked at a later stage, if the increased degradation will destroy mRNA even before it can arrive at this blocked later stage. In some cases (such as the mixed mechanism *F* on the plot, co-existence of complete Cap Inhibition and Decay), the kinetic signature of the mixed mechanism is indistinguishable from Decay.

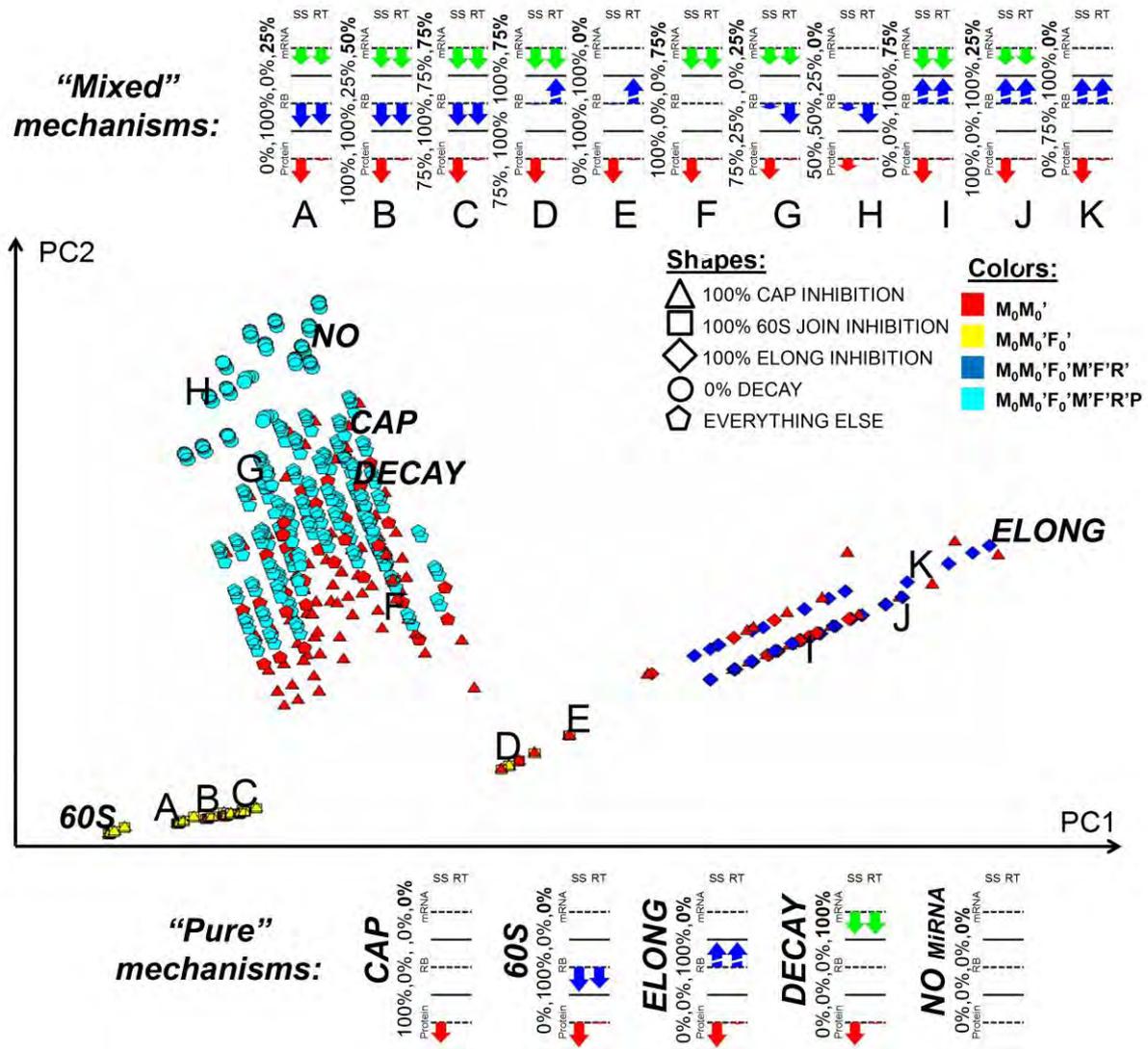

Figure 13.8. PCA plot for simulations for a fixed set of translation parameters and a variable mixed mechanisms. The mix includes (1) Cap Inhibition, (2) 60S unit joining inhibition, (3) Elongation inhibition and (4) Decay for the reference set of translation parameters and $k_b = 10^{-3}$. The plot represents a projection from a six-dimensional space of measurable quantities: relative changes in steady-state (*SS*) and relaxation time (*RT*) for three quantities: amount of mRNA (*mRNA*), number of ribosomes per mRNA (*RB*) and amount of protein (*Protein*). Each point represents a simulation made for a selected spectrum of strengths of four mechanisms, the colors distinguish the resulting dominant paths and the shapes distinguish spectrums when one of the mechanisms is dominating or when the mRNA decay is not affected by miRNA (circles). For example, red rectangle corresponds to the scenario when 60S unit joining is completely blocked if miRNA is bound, and $M_0M_0'$ dominant path is realized. Few points are annotated with signature diagrams visualizing the numerical values for the six variables. Four numbers on the left of each diagram show the strengths of four miRNA action mechanisms (cap inhibition, 60S unit joining inhibition, elongation inhibition and decay correspondingly). First two principal components explain 86% of data variation.

The kinetic signature *K* (mix of 60S Unit Joining Inhibition and Elongation Inhibition) is indistinguishable from the pure signature of Elongation Inhibition. The kinetic signature *H* (mix of three first mechanisms without Decay) reminds pure 60S Unit Joining Inhibition mechanism. Cases *F*, *K* and *H* are three examples of *kinetic signature masking* (or domination) of one mechanism by another.

In other cases the resulting kinetic signature of a mixed mechanism does not remind any signature of the four pure mechanisms: by contrast, certain superimposition of the kinetic signatures happens. Thus, the mixed mechanisms *A* (co-existence of complete 60S Unit Joining Inhibition and Decay) and *E* (co-existence of complete 60S Unit Joining Inhibition and Elongation inhibition) give the signature which looks like a superimposition of the kinetic signatures of the initial mechanisms. However, further addition of miRNA action mechanisms does not change the signature qualitatively. Thus, mix of all four mechanisms together (cases *B*, *C*) still looks like a mix of 60S Unit Joining Inhibition and Decay. Hence, one can say that in this case a superimposition of two kinetic signatures masks signatures of other mechanisms.

Interestingly, the kinetic signature in the mixed mechanism *J* (mix of Cap Inhibition, Elongation Inhibition and Decay) can be still interpreted as a mix of three signatures of the initial pure mechanisms. This is an example, when three mechanisms are superimposed and leave their "traces" in the final mix.

## 8.2. Fixed mixed mechanism of miRNA action and variable experimental or cellular context of translation

In the second computational experiment we fixed the strengths of the four mechanisms at 50%, i.e. we consider the mixed miRNA action mechanism characterized by the spectrum $S = \{0.5, 0.5, 0.5, 0, 0, 0, 0.5, 0, 0\}$. For the reference set of parameters and variable miRNA-mRNA binding constant (see Figure 13.9, top left), this mixed mechanism of miRNA action is manifested by a kinetic signature which can be attributed to the Decay mechanism of miRNA action (M7). However, for other parameter combinations the kinetic signature of this mechanism can look differently and expose features of other mechanisms (see below). The main message of the example shown in Figure 13.9 is that variation of the parameters of translation mechanism can significantly change the interpretation of the kinetic signature when several mechanisms of miRNA action co-exist.

We study the kinetic signatures of the mixed mechanism $S = \{0.5, 0.5, 0.5, 0, 0, 0, 0.5, 0, 0\}$ when the kinetic parameters of the normal translation are varied in very large intervals (five orders of magnitude). We varied four kinetic rates $k_d$, $k_b$, $k_{01}$, $k_2$, leaving $k_t$ and $k_p$ fixed at the reference values and putting $k_3 = k_2/6$ to provide constant average number of 6 ribosomes sitting on one mRNA. The parameters took the following range of values: $k_d \in \{10^{-3}, 10^{-4}, 10^{-5}, 10^{-6}, 10^{-7}\}$, $k_b, k_{01}, k_2 \in \{10^{-1}, 10^{-2}, 10^{-3}, 10^{-4}, 10^{-5}\}$ in all possible combinations. From these combinations those were excluded that violated the condition of efficient translation (not dominated by degradation) $k_d \ll k_{01}, k_2, k_3$. As a result, we have tried 440 different simulations for which we created kinetic signatures, characterized by 6 numbers, as previously. These signatures can be represented as a cloud of 440 points in the 6-dimensional space, which was projected on a 2D plane using the standard principal components analysis (PCA), see Figure 13.9. This figure represents a "portrait" of a mixed mechanism of miRNA action for all relevant parameter values, including the reference parameter values (points *RW*, *RM*, *RS*).

The two-dimensional distribution of kinetic signatures shown in Figure 13.9 shows two tendencies. Moving from top right to bottom left corner (from point *A* to point *E*) corresponds to increasing relative value of the miRNA binging constant, leading to more complete inhibition of protein synthesis. Moving from top left to bottom right corner (from point *F* to point *J*) corresponds to changing mainly the relaxation time of the ribosomal profile. There is a third degree of freedom not shown in the figure and associated with the third principal component which is almost completely corresponds to the change in the protein synthesis relaxation time. Thus, some points located closely on the plot (such as points *C* and *D*) are in fact separated along the third principal component and have very different protein synthesis relaxation time values.

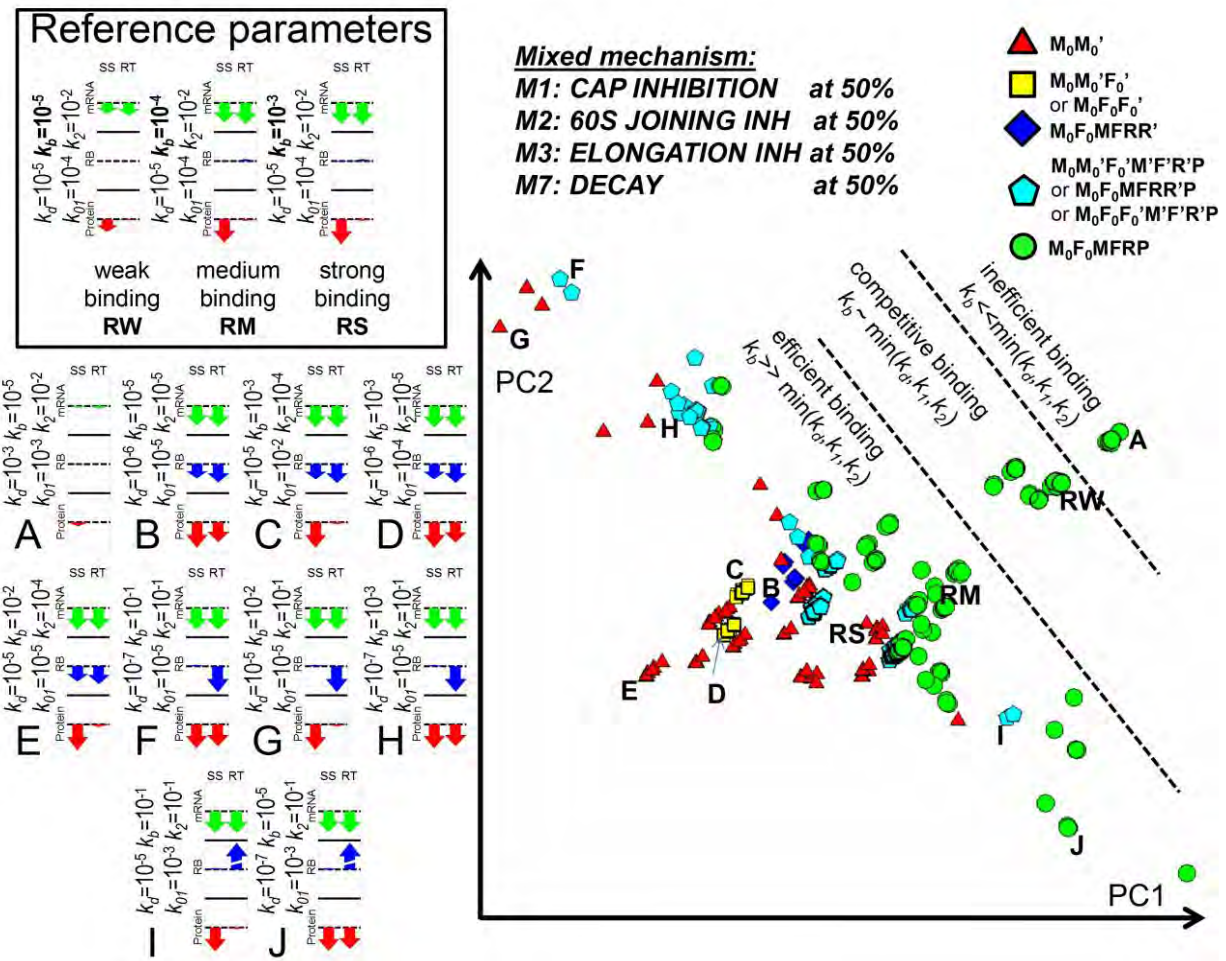

Figure 13.9. PCA plot (on the right) for simulations for one selected mixed mechanism (Cap Initiation Inhibition at 50%, 60S Unit Joining Inhibition at 50%, Elongation Inhibition at 50% and Decay at 50%) and a variable set of internal translation parameters. The plot represents a projection from six-dimensional space of measurable quantities: relative changes in steady-state (*SS*) and relaxation time (*RT*) for three quantities: amount of mRNA (*mRNA*), number of ribosomes per mRNA (*RB*) and amount of protein (*Protein*). Each point represents a simulation made for a combination of $k_d$, $k_b$, $k_{01}$ and $k_2$ parameter values, the color and shape distinguish the resulting dominant paths. Several points are annotated with signature diagrams visualizing the numerical values for the six variables (on the left). Cases *RW*, *RM* and *RS* (top left) show close to the reference set values for three different miRNA binding rates ($k_b$=10$^{-3}$,10$^{-4}$,10$^{-5}$). First two principal components explain 79% of data variation.

Several important conclusions can be made from this computational experiment, and the first one concerns the role of miRNA binding strength. Evidently, if $k_b$ is much smaller than the normal translation parameters $k_d$, $k_{01}$, $k_2$ then miRNA binding does not affect the dynamics significantly and the "normal" $M_0F_0MFRP$ dominant path is implemented (case *A*). In the case when the binding is significant but not very strong and comparable to $k_d$, $k_{01}$, $k_2$ parameters (competitive binding), the signature is masked by Decay-like pattern (case *RW*). The Decay mechanism masks all other mechanisms also in those combinations of parameters where $k_2$ is faster than $k_d$ by several (three) orders of magnitude (cases *RM* and *RS*). In this case, the ribosomal profile is not perturbed by miRNA.

The relaxation time of a protein changes in the signatures when the mRNA degradation rate becomes less than the degradation rate of the protein: $k_d \ll k_p = 5 \cdot 10^{-6}$ (see formula (15)). Notice that for the reference set of parameters the protein is assumed to be more stable than a transcript, and the only "pure" signature where the relaxation time of the protein is affected by miRNA is Cotranslational Protein Degradation. If the protein is less stable than a transcript then this might create confusion in interpreting the signatures and suggesting activation of this mechanism while it is not functional in reality.

The signatures *B*, *C*, *D* and *E* can be interpreted as a superimposition of 60S Unit Joining Inhibition with Decay, with possible role of Cap Inhibition. Elongation Inhibition mechanism leading to the increase of both RB steady state and relaxation time might be suspected in the signatures *I* and *J* even though the RB steady state does not change significantly (this can be attributed to a compensatory effect from mixing Elongation Inhibition with 60S Unit Joining). Signatures *F*, *G* and *H* suggest the role of 60S Unit Joining Inhibition (decreasing the ribosome profile relaxation time) which would be completely missed if one looks at the relative changes of the steady states only.

Finally, let us notice the special role of the $M_0M_0'$ dominant path which can produce kinetic signatures very similar to other dominant paths (this is true both for Figure 13.8 and Figure 13.9). Compare, for example, pairs of cases *F* and *G*, *D* and *E*. The dominant path $M_0M_0'$ requires relatively strong binding $k_b \gg k_{01}$ and relatively fast degradation or slow initiation, which can be expressed as condition on parameters $\dfrac{k_{01}'}{k_d'} = \dfrac{(1-s_1)k_{01}}{(1+9s_4)k_d} \ll 1$, where $s_1$ and $s_7$ are the strengths of the Cap Initiation Inhibition and Decay mechanisms respectively in the mixed mechanism. In the case when $s_1 = s_7 = 0.5$, this gives a condition $k_{01} \ll 9k_d$, i.e. that the normal cap initiation rate should not exceed the normal degradation rate by more than two orders of magnitude (100-fold). This condition is satisfied for the points *B*, *E*, *G* and the reference set of parameters on the plot in the Figure 13.9. On the other hand, it can be shown that $k_{01}$ is the least sensitive parameter affecting the relative changes of the steady states and relaxation times for the MT, RB and PR values (decreasing $k_{01}$ can affect only the steady state of the protein and not other values, see (14)). Hence, for many kinetic signatures, given relatively strong miRNA binding constant, there is a possibility to implement the $M_0M_0'$ dominant path by slowing down $k_{01}$ without a qualitative signature change. This non-intuitive conclusion can be verified experimentally.

## 9. Concluding remarks

MicroRNA mode of action is a highly controversial topic. Here, we used mathematical modelling and found that each of the suggested mechanisms has a specific signature (the predicted dynamics of 3 measurable variables of the translational process, namely, the time course of accumulation of protein and mRNA, and of ribososmal loading on the mRNA). These signatures provide a new tool for discriminating between distinct mechanisms. We thus propose the **concept of a characteristic kinetic signature for miRNA modes of action.**

In addition, an essential conclusion of our analysis is that miRNA action will impact the final kinetic output only if it targets a **sensitive parameter of the system**.

The hypothesis that microRNA action can have a visible impact on protein output only if it affects the rate-limiting step has already been suggested in [62] for inhibition of translational initiation. However, the notion of rate-limiting step becomes non-trivial when we consider complex networks (more complex than a linear chain or a cycle of monomolecular reactions). The mathematical model that we present here confirms the conclusions from [62], and extends them to all steps of microRNA action. The mathematical approach we have developed for analysis of this complex system uses the notion of dominant dynamical system, itself a generalization of the rate-limiting step concept to complex networks [22,23,90]**.**

In accordance with the general theory of dynamical limitation [22], we can take into account not only the steady-state rates of protein synthesis but also its relaxation time. For example, for a linear chain of reactions, the steady-state rate depends on the slowest kinetic rate parameter (rate-limiting step), whereas the relaxation time of the system depends on the second slowest kinetic rate parameter.

The analysis of our results allowed us to suggest a unifying theory for miRNA modes of action: all proposed modes of action operate simultaneously, and the *apparent* mechanism that will be detected depends on a set of sensitive intrinsic parameters of the individual target mRNA under study. This hypothesis would explain the following set of observations: 1) that the same microRNA apparently uses distinct mechanisms on different targets (e.g. for let7: [11,54,55,69,83,]; for CXCR4: [29,67,84,85]; for miR16: [28,37]; for miR122: [35]); 2) that microRNA's mode of action depends on the promoter under which the target mRNA is transcribed [42]; and 3) that the status of the cell affects the final observable mode of miRNAs action [8,47,81]. Moreover, the possibility of coexistence of two or several mechanisms has already been discussed and proven in the literature [15,16,19,47,69,81,87,91].

As already stated, our modelling results lead us to propose that, in individual biological systems, the relative abundance and/or activity of some set of intrinsic factors determines the apparent inhibition mechanism that will be detected. These factors are *not related to the miRNA pathways*, but intrinsically determine the **sensitive parameters of the system**. Indeed, RNA-binding proteins not related to the miRNA pathway have been shown to have a strong influence on the final outcome of miRNA regulation [56,59,74,88].

A body of studies underscore the importance of intrinsic parameters of mRNAs. Revisiting theses studies in the framework of our model provides an explanation for most of the discrepancies in the literature. Thus, in most of the studies showing initiation inhibition, *in vitro* transcribed mRNAs (transfected into cells or studied directly *in vitro*) were used. In contrast, almost all data supporting elongation inhibition were obtaining in living cells, and thus with physiologically modified target mRNAs [29,40,55,69,80,85], with only one and very specific exception [53]. Similarly, most of the studies showing IRES-driven mRNAs as being refractory to microRNAs were carried out either *in vitro* [55] or using *in vitro* transcribed mRNAs transfected into cells [29,40.69], whereas the studies showing IRES-driven mRNAs to be repressed by miRNAs were carried out with mRNAs transcribed *in situ*, inside cells [37,67]. In all these cases, the difference might come from the status of the target mRNA, rather than from any putative or actual differences in the microRNA machinery.

Another example is the influence of splicing marks attached to mRNAs *in vivo*. The process of mRNA splicing leaves protein marks on mRNAs, which promotes the first round of translation at the initiation step [46,59]. These marks are dissociated during the first round of translation. Splicing marks, by increasing the initial initiations, would lead to higher initiation rates on intron-containing mRNAs [36]. Elongation would thus become a limiting step. In contrast, *in vitro* transcribed mRNAs lack splicing marks, resulting in a decreased initiation rate, which becomes limiting. Moreover, under *in vitro* conditions, initiation is highly dependent on the concentration of initiation factors, providing another possible explanation for discrepancies between *in vitro* studies.

Another example is the dependence of miRNA effects on codon usage. MicroRNA action has been reported to act on initiation steps when codon usage is optimized for human translation [40,68], whereas, with non-optimized codons, microRNA was found to act on elongation [25,53,63,67]. This again might have something to do with different rates of elongation, elongation rates being, or not, among the set of limiting (sensitive) parameters for a given mRNA.

Yet another example is the dependence of microRNA mode of action on the experimental procedure for transfection of the mRNA [53]. The transfection procedure is likely to influence the association of the target mRNA with mRNA-binding proteins, which, in turn, changes the sensitive parameters of the system, and hence the final outcome of microRNA action.

All these and some other data clearly support the idea that the observed mode of action of a microRNA depends upon interplay between the intrinsic rates of the different steps of mRNA translation.

In summary, our results provide a mathematical tool to discriminate between different miRNA modes of action. Moreover, we propose a unifying model in which the observed mode of action of a particular miRNA is dictated by the relationships among the intrinsic parameters of its target mRNA. We anticipate that the tool we have developed will promote better analysis of experimental data, and that our model will permit a better understanding of microRNA action. Most importantly, our hypothesis would explain most of the discrepancies in the corresponding literature.

**Acknowledgements**

This work was supported by a grant from the European Commission Sixth Framework Programme (Integrated Project SIROCCO, contract number LSHG-CT-2006-037900) to AHB, and from the Agence Nationale de la Recherche (project ANR-08-SYSC-003 CALAMAR) and from the Projet Incitatif Collaboratif "Bioinformatics and Biostatistics of Cancer" to Institut Curie. AZ is a member of the team "Systems Biology of Cancer", labeled by the Ligue Nationale Contre le Cancer.